\documentclass[final]{elsart}

\usepackage[dvips]{graphicx}
\usepackage{amssymb}

\begin{document}

\begin{frontmatter}

\title{A New Numerical Technique to Determine Primary Cosmic Ray Composition
in the Ankle Region}

\author[UNAM]{A. D. Supanitsky\thanksref{Tandar}\corauthref{cor}},
\author[UNAM]{G. Medina-Tanco}, and
\address[UNAM]{Instituto de Ciencias Nucleares, UNAM, Circuito Exteriror S/N, Ciudad Universitaria,
M\'exico D. F. 04510, M\'exico.}
\author[Tandar]{A. Etchegoyen\thanksref{CONICET}}
\address[Tandar]{Departamento de F\'isica, Comisi\'on Nacional de Energ\'ia At\'omica, Av. Gral. Paz 1499, 
Buenos Aires, Argentina.}
\thanks[CONICET]{Member of Carrera del Investigador Cient\'ifico, CONICET, Argentina.}
\corauth[cor]{Corresponding author. Present Address: Instituto de Ciencias Nucleares, UNAM, Circuito Exteriror S/N,
Ciudad Universitaria, M\'exico D. F. 04510, M\'exico. E-mail: supanitsky@nucleares.unam.mx.}

\begin{abstract}
In this paper we introduce a new multiparametric technique that attempts to tackle simultaneously the problems 
of composition determination and hadronic interaction uncertainty. Employing simulations of a real world detector 
under its planned operational conditions, and disregarding systematics, we can asses that the present technique 
should be able to determine the composition of a binary mixture of p and Fe with a statistical confidence of few 
percent, in a way that is independent of the assumed hadronic interaction model. Moreover, the combination of real 
data with the tools developed and presented here should give an indication of the reliability of the various hadronic 
interaction models in current use in the area. We center our study in the region of the ankle, where composition 
carries critical astrophysical information, and use two main parameters: the number of muons at $600$ m from the 
shower axis and the depth of the shower maximum obtained from the hybrid operation of the planned muon counters and 
high elevation fluorescence telescopes of the AMIGA and HEAT Auger enhancements.
\end{abstract}

\end{frontmatter}

\section{Introduction}
\label{Int}

The cosmic ray energy spectrum, in the high energy region 
above $10^{17}$ eV, presents two main features observed 
by several experiments, the second knee, observed at 
$4 \times 10^{17}$ eV \cite{Nagano:84,Abu:01,Pravdin:03,HiRes:04} 
and the ankle. There is evidence of a fourth feature situated at 
the highest observed energies, consistent with the so-called GZK 
suppression \cite{GZKAuger,GZKHiRes}, which would be caused by 
the interaction of the ultra-high energy protons with the photons 
of the cosmic microwave background radiation (CMBR) 
\cite{Greisen:66,Zatsepin:66}. For the case of heavier nuclei a 
similar effect is expected because of their interaction with 
photons from the infrared and microwave backgrounds 
\cite{Stecker:69}.

The origin of the second knee is still unclear. It has been
interpreted as the end of the efficiency of the acceleration 
in Galactic supernova remnant shock waves \cite{CinziaICRC:07}, 
a change in the diffusion regime in our galaxy 
\cite{Hoerandel:03,Candia:02} or even the transition between 
the Galactic and extragalactic components of the cosmic rays 
\cite{Berezinsky:04}.

The ankle is a broader feature. It has been observed by Fly's Eye
\cite{Abu:01}, Haverah Park \cite{Ave:01}, Yakutsk
\cite{Pravdin:03}, HiRes \cite{HiRes:04} and Auger \cite{GZKAuger}
in Hybrid mode at approximately the same energy, $\sim 3\times
10^{18}$ eV. The origin of the ankle is also unknown. It can be 
interpreted as the transition between the Galactic and extragalactic 
components \cite{Allard:05} or the result of pair production by 
extragalactic protons after the interaction with photons of the 
CMBR during propagation \cite{Berezinsky:88,Berezinsky:05}.

There are three main models of the Galactic-extragalactic
transition. The first is the mixed composition model \cite{Allard:05}, 
in which extragalactic sources inject a spectrum of masses 
similar to lower energy Galactic cosmic rays and for which the 
transition takes place at the ankle. The second is the ankle model 
\cite{Wibig:05}, a two-component transition from Galactic iron nuclei 
to extragalactic protons at the ankle energy. The third is the dip
model \cite{Berezinsky:04}, in which the ankle is due to pair production 
of extragalactic protons that interact with the photons of the CMBR (in 
this scenario the transition occurs at the second knee). 

In order to rule out or substantiate any of those models, additional information 
is necessary besides the energy spectrum shape and absolute intensity. Detailed 
measurements of the composition as a function of energy, while not sufficient, 
would be extremely valuable to break the present degeneracy among competing models 
for the Galactic-extragalactic transition \cite{MedinaTanco:07,GMT_EMA:06}. 
Furthermore, this kind of information could help to determine what the highest 
energy accelerators in the Galaxy are and provide indicators of the kind and 
level of magnetohydrodynamic turbulence present in the intergalactic medium 
traversed by the lowest energy cosmic ray particles \cite{CinziaICRC:07}.

Several experiments have measured the cosmic ray composition in
the region where the transition takes place. Nevertheless,
large discrepancies exist between different experiments and experimental
techniques \cite{Dova:05}. One of the main reasons behind the plurality of
sometimes contradicting results is that composition is determined by 
comparing experimental data with numerical shower simulations. These 
simulations include models for the relevant hadronic interactions which are  
extrapolations, over several orders of magnitude in center of mass, of accelerator 
data to cosmic ray energies. This is a source of considerable uncertainty
which is confirmed, to a certain extent, by the fact that there is experimental 
evidence of a deficit in muon content of simulated showers with respect to real 
data \cite{Engel:07}.

In this paper we introduce a new statistical method to test the
compatibility of the high energy hadronic interaction models
and real data. In fact, the hadronic interaction model is assumed
to be the main systematic uncertainty in the present analysis. Throughout
our analysis, we consider QGSJET-II \cite{QGIIa,QGIIb} and Sibyll 2.1
\cite{Sib2.1}. Our method allows us to verify whether the experimental data
are compatible with the hadronic models under consideration and, if so, to
estimate the composition. 

Although this new technique is of general applicability, we study its potential 
in the context of AMIGA (Auger Muons and Infill for the Ground Array) 
\cite{Etchegoyen:07} and HEAT (High Elevation Auger Telescopes) \cite{Klages:07}, 
the lower energy extensions of the Southern Pierre Auger Observatory. These two 
enhancements will extend the energy range down to $10^{17}$ eV, encompassing the 
second knee and ankle region where the Galactic-extragalactic transition takes 
place. 

The most relevant mass sensitive parameters that will be obtained from AMIGA 
and HEAT are the number of muons at 600 m from the shower core, $N_{\mu}(600)$,
and the atmospheric depth of maximum shower development, $X_{\mathrm{max}}$, 
respectively. Consequently, we develop our statistical technique by using mainly 
this pair of parameters. Combinations of $N_{\mu}(600)$ with parameters from the 
Cherenkov detectors like the slope of the lateral distribution function, rise-time 
of the signals and curvature radius are also studied. 

Note that our technique should also be applicable to Telescope Array and its low energy 
extension \cite{Martens:07,Belz:07}, which have hybrid capabilities in the region of the 
ankle and also plan to include muon detectors \cite{Belz:07,TaleCollab}. The quoted expected 
error for $X_{\mathrm{max}}$ is $\sim 20$ g cm$^{-2}$ \cite{TaleCollab}, which is comparable 
to the value estimated for HEAT. 

$N_{\mu}(600)$ is nearly linearly dependent on energy; therefore, its use as a composition 
estimator requires, ideally, an independent determination of the shower energy. This is not a 
problem in the case of the Auger enhancements, where the same strategy of energy calibration 
as with the baseline design can be used. Hybrid events from the AMIGA-HEAT detector provide a 
calibration for $S(r_{0})$, the lateral distribution function value at a fixed distance 
$r_{0} \sim 600$ m, as measured by the AMIGA surface array of water Cherenkov detectors. The 
same procedure should be, in principle, applicable to other detectors with hybrid capability 
like Telescope Array. This is so because $N_{\mu}(600)$ and energy are correlated through $S_{600}$ 
and the latter also receives an important contribution from the electromagnetic lateral distribution 
function, while $N_{\mu}(600)$ is directly used as a composition parameter. Therefore, at energies 
of few $10^{18}$ eV or smaller, under the assumption of a binary mixture of proton and iron, the 
separation in $N_{\mu}(600)$ due to composition alone at fixed energy is larger than the separation 
in $S_{600}$ due to composition at the same energy. This can be clearly seen, for example, in figures 
12.a and 12.b of Ref. \cite{SupaRec:08}, where the merit factors (or discrimination power) for 
different composition indicators are shown as a function of energy for two zenith angles. The merit 
factor is, basically, the separation between the distribution functions of each parameter at a given 
energy normalized by the combined dispersion of both distributions. It can be seen that, at a fixed 
energy, the separation between the distributions of $N_{\mu}(600)$ for proton and iron, the main 
composition indicator, is much larger than the separation between the corresponding distributions 
of $S_{600}$, the energy estimator. Consequently, the fact that $N_{\mu}(600)$ is correlated 
simultaneously to energy and composition does not inhibit its use as a composition parameter.

Section \ref{CompoDet} shows our main results. In particular, it is shown there that,
by working on a plane defined by two parameters derived from $X_{\mathrm{max}}$ and
$N_{\mu}$ data, an estimation of the cosmic ray composition can be obtained that 
is reasonably independent of the uncertainties related to the underlying hadronic 
interaction model. Furthermore, we demonstrate that rather small samples of events at 
a given reconstructed energy, compatible with the level of statistic expected from 
detectors currently under construction that will operate in the ankle region, are enough 
to this end. Additionally to the determination of composition, given two possible 
interaction models and an observed data set, the technique can be used to assess the 
compatibility of these models with the experimental data.

\section{Abundance Estimator}
\label{AbEst}

In order to develop a statistical method to infer the composition of the cosmic rays
(i.e. the abundance of a given primary type), let us consider two possible types of 
primaries, $A = a, b$, and samples of size $N = N_{a}+N_{b}$, where $N_a$ and $N_b$ are 
the number of events corresponding to type $a$ and $b$, respectively. From each event of 
an individual sample it is possible to extract several observable parameters sensitive to 
the primary mass. Therefore, for a given mass sensitive parameter $q$ we define,
\begin{equation}
\xi_{q} \equiv \frac{1}{N}\ \sum^{N}_{i=1} P_{a}(q_{i}) = \frac{1}{N}\ \left[ \sum^{N_{a}}_{i=1}%
P_{a}(q_{i}^{a}) + \sum^{N_{b}}_{i=1} P_{a}(q_{i}^{b}) \right],
\label{XiDef}
\end{equation}
where $q_{i}^{A}$ are $N_{A}$ random variables distributed as $f_{A}(q)$ and
\begin{equation}
P_{a}(q) = \frac{f_{a}(q)}{f_{a}(q)+f_{b}(q)},
\label{Pa}
\end{equation}
which is the probability that an event is of type $a$ for a given $q$, assuming no prior knowledge
of the primary type. Note that we restrict our analysis to the case in which the cosmic rays are
the superposition of two components.

The new statistic, so defined, is an estimator of the abundance of the primary of type $a$.
$\xi_{q}$ is a random variable because it is a function of $N$ random variables,
$\xi_{q} = \xi_{q}(q_{1}^{a} \ldots q_{N_{a}}^{a},q_{1}^{b} \ldots q_{N_{b}}^{b})$. Therefore, 
its mean value as a function of the composition is given by,
\begin{eqnarray}
\langle \xi_{q} \rangle (c_{a}) &=& \int dq_{1}^{a} \ldots dq_{N_{a}}^{a} dq_{1}^{b} \ldots dq_{N_{b}}^{b}%
\ \xi_{q}(q_{1}^{a} \ldots q_{N_{a}}^{a},q_{1}^{b} \ldots q_{N_{b}}^{b}) \times  \nonumber \\
&& f_{a}(q_{1}^{a}) \cdots f_{a}(q_{N_{a}}^{a})\ f_{b}(q_{1}^{b}) \cdots f_{b}(q_{N_{b}}^{b}),
\label{MVXiInt}
\end{eqnarray}
where $c_{a} = N_{a}/(N_{a}+N_{b})$ is the composition (or the abundance) corresponding to the primary $a$ 
for the sample of $N$ events. Noting that $f_{A}(q)$ are probability density functions and, therefore, that 
$\int dq \ f_{a}(q) \equiv 1$, the dependence of $\langle \xi_{q} \rangle$ on $c_{a}$ can be explicitly seen 
after integrating Eq. \ref{MVXiInt},   
\begin{equation}
\langle \xi_{q} \rangle (c_{a}) = m\ c_{a} + d,
\label{MVXi}
\end{equation}
where,
\begin{eqnarray}
\label{m}
m &=& \int dq P_{a}(q) (f_{a}(q)-f_{b}(q)), \\
\label{b}
d &=& \int dq P_{a}(q) f_{b}(q).
\end{eqnarray}

Equation (\ref{MVXi}) shows that the mean value of $\xi_{q}$ increases linearly with the composition of 
the samples and from equations (\ref{m},\ref{b}) we see that as the overlap between the distributions $f_{a}(q)$ 
and $f_{b}(q)$ decreases, $\langle \xi_{q} \rangle (c_{a})$ tends to the identity function. We also see that 
$\langle \xi_{q} \rangle (1/2) = 1/2$, independent of the shape of the distributions $f_{a}(q)$ and
$f_{b}(q)$.

The variance of $\xi_{q}$ is given by,
\begin{equation}
Var[ \xi_{q} ] (c_{a}) = \frac{1}{N}\left[ c_{a} \left( \sigma^{2}_{a}[P_{a}(q)] - \sigma^{2}_{b}[P_{a}(q)]%
\right)+\sigma^{2}_{b}[P_{a}(q)] \right],
\label{VarXi}
\end{equation}
where,
\begin{equation}
\sigma^{2}_{A}[P_{a}(q)] = \int dq P_{a}^{2}(q) f_{A}(q) - \left( \int dq P_{a}(q) f_{A}(q) %
\right)^{2}.
\label{SigPa}
\end{equation}
Therefore, the variance of $\xi_{q}$ also increases linearly with $c_a$. Note that it is
proportional to $N^{-1}$, i.e., it approaches zero as the size of the sample approaches infinity.

The composition estimator $\xi_{q}$ is defined for each parameter $q$, and therefore a 
covariance can be calculated for any two of such estimators, say $q_{\mu}$ and $q_{\nu}$. Let 
$f_{A}(q_{\mu}, q_{\nu})$ be the distribution function of these variables and
\begin{eqnarray}
f_{\mu A}(q_{\mu}) &=& \int dq_{\nu}\ f_{A}(q_{\mu}, q_{\nu}),  \label{faa} \\
f_{\nu A}(q_{\nu}) &=& \int dq_{\mu}\ f_{A}(q_{\mu}, q_{\nu}), \\
P_{\mu a}(q) &=& \frac{f_{\mu a}(q)}{f_{\mu a}(q)+f_{\mu b}(q)}, \\
P_{\nu a}(q) &=& \frac{f_{\nu a}(q)}{f_{\nu a}(q)+f_{\nu b}(q)}.
\end{eqnarray}
The covariance between $\xi_{q_{\mu}}$ and $\xi_{q_{\nu}}$ is given by,
\begin{eqnarray}
Cov(\xi_{q_{\mu}},\xi_{q_{\nu}}) &=& \frac{1}{N} \Big[ \ c_{a} \int dq_{\mu} dq_{\nu}\ %
P_{\mu a}(q_{\mu}) P_{\nu a}(q_{\nu})\ (f_{a}(q_{\mu}, q_{\nu})-  \nonumber \\
&& f_{\mu a}(q_{\mu}) f_{\nu a}(q_{\nu}))+(1-c_{a}) \int dq_{\mu} dq_{\nu}\ P_{\mu a}(q_{\mu})
P_{\nu a}(q_{\nu})\times \nonumber \\
&& (f_{b}(q_{\mu}, q_{\nu})- f_{\mu b}(q_{\mu}) f_{\nu b}(q_{\nu})) \Big].  \nonumber
\label{Cov}\\
\end{eqnarray}
From Eq. (\ref{Cov}) we can see that $Cov(\xi_{q_{\mu}},\xi_{q_{\nu}})$ is also a linear function of 
$c_a$, and a sufficient condition to be zero is that the variables $q_{\mu}^{A}$ and $q_{\nu}^{A}$
are independent.

The parameter $\xi_{q}$ is the sum of $N$ random variables, therefore, for large enough values of $N$, 
this variable follows a Gaussian distribution because of the central limit theorem.

\begin{figure}[!bt]
\begin{center}
\includegraphics[width=13cm]{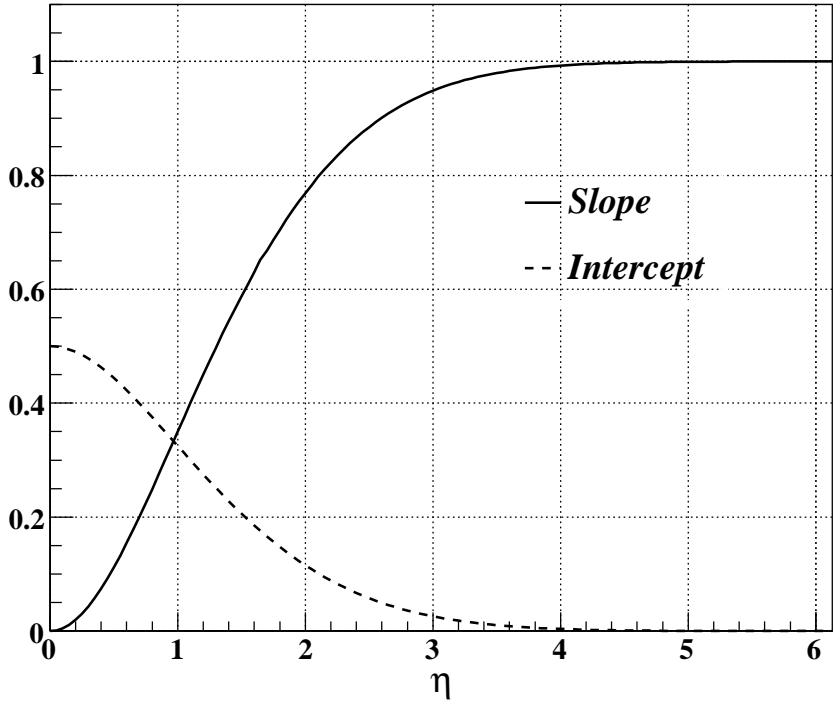}
\caption{Slope and intercept of $\langle \xi \rangle (c_{a})$ as a function of $\eta$ for two 
Gaussian distributions with $\sigma = 1$. \label{Example}}
\end{center}
\end{figure}

As a simple example of Eq. (\ref{MVXi}), let us consider two Gaussian distribution functions of mean 
values $+x$ and $-x$ and $\sigma = 1$. Under this assumption, the parameter
$
\eta=\left| \langle q_{1} \rangle - \langle q_{2} \rangle \right| / [\sigma^{2}(q_{1}) +%
\sigma^{2}(q_{2})]^{1/2}
$,
which measures the discrimination power of $q$, can be easily calculated and takes the values 
$\eta = x/\sqrt{2}$. Fig. \ref{Example} shows the corresponding slope and the intercept of 
$\langle \xi \rangle (c_a)$ as a function of $\eta$. We see that, as $\eta$ increases, the
intercept approaches zero and the slope approaches one, i.e., as expected, 
$\langle \xi \rangle (c_{a})$ tends to the identity function.

\section{Simulations and Composition Analysis}
\label{AnSim}

\subsection{Shower and detector simulations}
\label{Sim}

In order to perform composition analyses around $E = 1$ EeV (EeV $= 10^{18}$ eV), including the effect of the 
energy uncertainty, we generated a library of atmospheric showers by using the AIRES package version 2.8.2 
\cite{AIRES}. We used a relative thinning of $0.6$ and statistical weight factor 0.2 (see Ref. \cite{AIRES} 
for details). For simplicity, and to a good approximation, we assume that the showers follow a power law energy 
spectrum with spectral index\footnote{The impact of the exact value of $\gamma$ in our results is negligible.} 
$\gamma=-2.7$ in the energy interval $[0.6, 2]$ EeV. We generated eight sets of 166 showers each corresponding 
to two kinds of primaries, protons and iron nuclei, two values of zenith angle, $\theta = 30^\circ$ and 
$\theta = 45^\circ$, and two different models of high energy hadronic interactions, QGSJET-II and Sibyll 2.1.

Each of these showers was used to generate events of the AMIGA-HEAT detector system. 
The simulation of the AMIGA detectors was performed with a dedicated package described in Ref. \cite{SupaRec:08}. 
Each shower was used 40 times by distributing impact points uniformly in the 750 m-array. We assumed 30 m$^2$ muon 
detectors segmented in 192 cells and buried 2.5 m underground \cite{SupaRec:08}. We used a time binning of 20 ns 
and an efficiency of each segment equal to one. The shower arrival directions and core positions were reconstructed 
with the standard CDAS package Er-v3r4 \cite{CDAS} specifically developed to reconstruct the Cherenkov detector 
information in Auger. To reconstruct the muon lateral distribution function, we used the method introduced in 
Ref. \cite{SupaRec:08}.

The effects of the response of the HEAT telescopes and the reconstruction procedure for the longitudinal
profile in the determination of $X_{\mathrm{max}}$ were included using the approach of Ref. \cite{SupaRec:08}. 
For each simulated AMIGA event, corresponding to a given shower, we obtained a reconstructed $X_{\mathrm{max}}$ 
by taking a random value from a Gaussian distribution of mean value equal to the one calculated internally in 
AIRES (obtained by fitting the longitudinal profile with a Gaisser-Hillas function) and $\sigma$, for the 
corresponding energy, obtained from the  interpolation of the simulated data given in Ref. \cite{Klages:07}. 
Note that, despite the fact that the fluctuations in $X_{\mathrm{max}}$ have been included in a realistic way, 
the same cannot be said of the possible correlations between the fluctuations in fluorescence and surface 
parameters for hybrid events. Nevertheless, we expect that such correlations could be dealt with by proper 
quality cuts without affecting our main conclusions.

In short, for each simulated event, we obtained all the parameters with the corresponding fluctuations of the 
Cherenkov, muon and fluorescence detectors. We obtained 8 sets of $N_{0} \cong 166\times 40=6640$ events each 
(depending on the reconstruction efficiency). To refer to each set, we will use the notation $S(\theta,A,h)$ 
where $\theta=30^\circ, 45^\circ$, $A=$Proton, Iron and $h=$QGSJET-II, Sibyll 2.1.

\subsection{Probability density function estimation}
\label{DenEst}

In order to calculate $\xi$ (see Eq. (\ref{XiDef})), we need the distribution functions, $f_{A}(q)$, of 
the different parameters sensitive to the primary mass considered, including the effects of the detectors and 
reconstruction methods. We use the non-parametric method of kernel superposition 
\cite{Silvermann:86,Scott:92,Fadda:98,Merritt:94} as an estimate of these probability density 
functions from the simulated data. 

We assume a Gaussian error of $25\%$ in order to include the uncertainty in the determination of the primary 
energy, which has an important effect in the discrimination power of $N_\mu(600)$. Therefore, we consider samples 
of events obtained in the following way: for each simulated event, belonging to a given set $S(\theta,A,h)$, of 
real energy $E$, we estimate the reconstructed energy by taking a random value from a Gaussian distribution of 
mean $E$ and $\sigma = 0.25 \times E$. Since the bias introduced by the rapid fall of the spectrum shifts the 
center of the bin from $1$ EeV to $1.14$ EeV (see Appendix B), if the ``reconstructed'' energy falls in an energy 
interval of width $0.5$ EeV centered at $1.14$ EeV, the event is added to a new sample, $S_{\Pi_{r}}(\theta,A,h)$. 
We repeat this procedure $10$ times for each set $S(\theta,A,h)$. Therefore, we obtain $10$ samples 
$S_{\Pi_{r}}^i(\theta,A,h)$ with $i=1 \ldots 10$.

The parameters used in the analysis are defined as follows: (i) $N_{\mu}(600)$ is the number of muons at 600 m 
from the shower core, which is estimated from a fit to the lateral distribution of the number of muons measured by 
the AMIGA muon counters \cite{SupaRec:08}, (ii) $R$ is the curvature radius of the shower front, (iii) $\beta$ is the 
slope of the lateral distribution function of the signal in the water Cherenkov detectors and (iv) $t_{1/2}$ is a 
parameter constructed with the rise-time of the signal in a selected subset of the triggered water Cherenkov 
detectors,
\begin{equation}
t_{1/2} = \frac{1}{N_{T}} \sum^{N_{T}}_{i=1} (t^{i}_{50}-t^{i}_{10}) \times \left(%
\frac{400\ \textrm{m}}{r_{i}} \right)^{2},
\label{RiseTimeDef}
\end{equation}
where $N_{T}$ is the number of stations with signal greater than $10$ VEM (Vertical Equivalent Muon\footnote{The 
signal deposited in a water Cherenkov tank when fully traversed by a muon vertically impinging in the center 
of the tank \cite{Bertou:06}.}), $t^{i}_{10}$ and $t^{i}_{50}$ are the times at which $10\%$ and $50\%$ of the total 
signal is collected, respectively, and $r_{i}$ is the distance from the $i$-th station to the shower axis. Only 
stations at a distance to the shower axis greater than $400$ m are included in Eq. (\ref{RiseTimeDef}).

For each obtained sample $S_{\Pi_{r}}^i(\theta,A,h)$ we calculate the probability density estimate corresponding 
to four pairs of parameters: $(N_{\mu}(600),X_{\mathrm{max}})$, $(N_{\mu}(600),\beta)$, $(N_{\mu}(600),t_{1/2})$ 
and $(N_{\mu}(600),R)$. 

Better estimates of the density functions are obtained using the adaptive bandwidth method introduced by B. Silverman
\cite{Silvermann:86}. We perform a first estimation of each density function by using a Gaussian kernel with a 
fixed smoothing parameter,
\begin{equation}
\hat{f}_{0}(\vec{x}) = \frac{1}{N \sqrt{|\mathbf{V}|}\ 2 \pi\ h_{0}^2} \sum_{i=1}^{N}\ %
\exp\left[ -\frac{(\vec{x}-\vec{x}_i)^T \mathbf{V}^{-1} (\vec{x}-\vec{x}_i)}{2 h^2_{0}} \right],
\label{fest0}
\end{equation}
where $\vec{x}$ is a two-dimensional vector of one of the pairs of the parameters considered, $N$ is the size of the 
sample, $\mathbf{V}$ is the covariance matrix of the data sample and $h_{0} = 1.06 \times N^{-1/6}$ is the smoothing 
parameter corresponding to Gaussian samples. The latter is used very often in the literature \cite{Knuteson:02} because 
it gives very good estimates even for non Gaussian samples. From the density estimate obtained by using Eq. (\ref{fest0}) 
we calculate the parameters,
\begin{equation}
\lambda_{i} = \left[ \frac{\hat{f}_{0}(\vec{x}_{i})}{\left( \prod_{j=1}^{N} \hat{f}_{0}(\vec{x}_{j})%
\right)^{1/N} } \right]^{-1/2},
\label{lamda}
\end{equation}
and then we obtain the density estimate from,
\begin{equation}
\hat{f}(\vec{x}) = \frac{1}{N \sqrt{|\mathbf{V}|}\ 2 \pi} \sum_{i=1}^{N}\ \frac{1}{h^2_{i}}%
\exp\left[ -\frac{(\vec{x}-\vec{x}_i)^T \mathbf{V}^{-1} (\vec{x}-\vec{x}_i)}{2 h^2_{i}} \right],
\label{fest}
\end{equation}
where $h_{i} = h_{0} \ \lambda_{i}$.

\begin{figure}[!bt]
\begin{center}
\includegraphics[width=13cm]{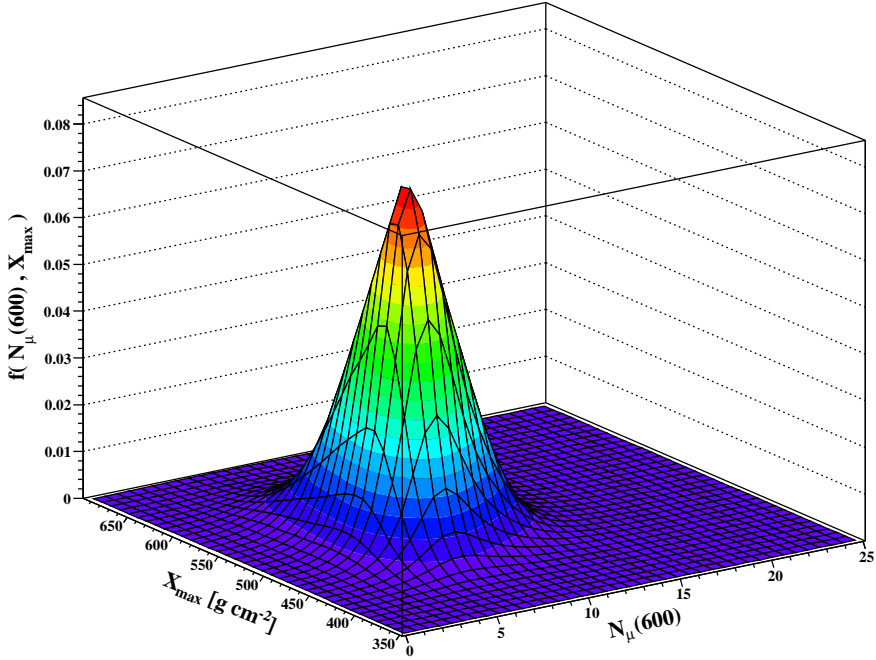}
\caption{Average distribution function for the parameters $X_{\mathrm{max}}$ and $N_{\mu}(600)$ corresponding to protons 
of $\theta=45^{\circ}$ and QGSJET-II as the high energy interaction model.
\label{ProbPr45deg}}
\end{center}
\end{figure}

Fig. \ref{ProbPr45deg} shows the average over the 110 density estimates corresponding to the parameters
$X_{\mathrm{max}}$ and $N_{\mu}(600)$ for protons, $\theta=45^{\circ}$ and QGSJET-II as the hadronic interaction
model.

Every estimate $\hat{f}(\vec{x})$ has an associated uncertainty because it is constructed from a finite set 
of $N$ elements. To take into account this uncertainty in the composition determination we use the smoothed 
bootstrap technique \cite{Silvermann:86}. For each $\hat{f}(\vec{x})$ we obtain 10 different samples, of 
the same size used to obtain $\hat{f}(\vec{x})$, by selecting random two-dimensional vectors from it. To 
every bootstrap sample we perform the same procedure done to the the original sample used to obtain 
$\hat{f}(\vec{x})$. Therefore, for every primary type, zenith angle, hadronic model and pair of parameters 
we obtain 110 estimates of the corresponding distribution function. Note that the smoothed bootstrap technique 
just allows us to estimate the variance of each $\hat{f}(\vec{x})$ but not the bias 
($Bias[\hat{f}(\vec{x})] = E[\hat{f}(\vec{x})]-f(\vec{x})$), which we assume is of negligible importance 
because we are using adaptive smoothed parameters.

\begin{figure}[!bt]
\begin{center}
\includegraphics[width=6.8cm]{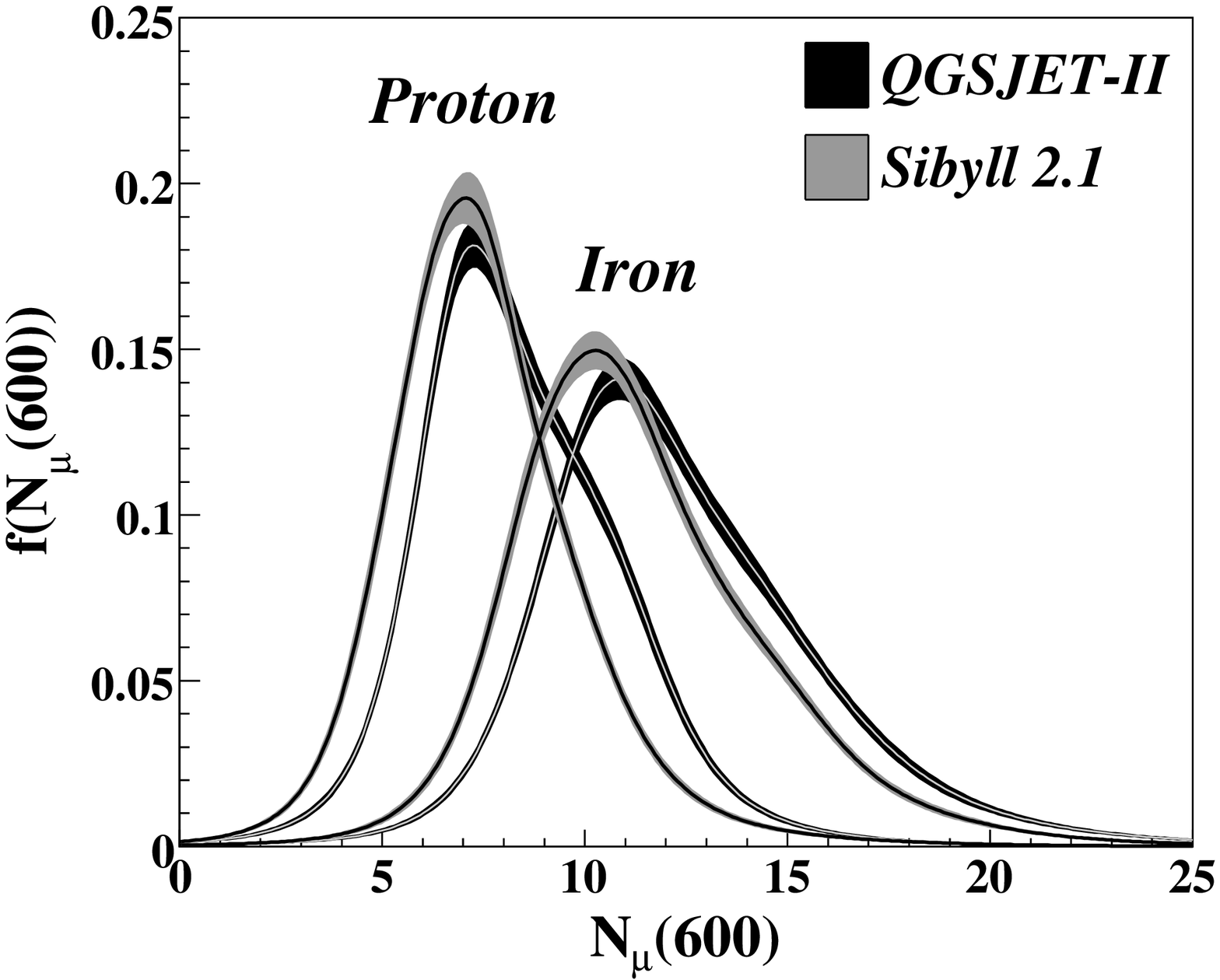}
\includegraphics[width=6.8cm]{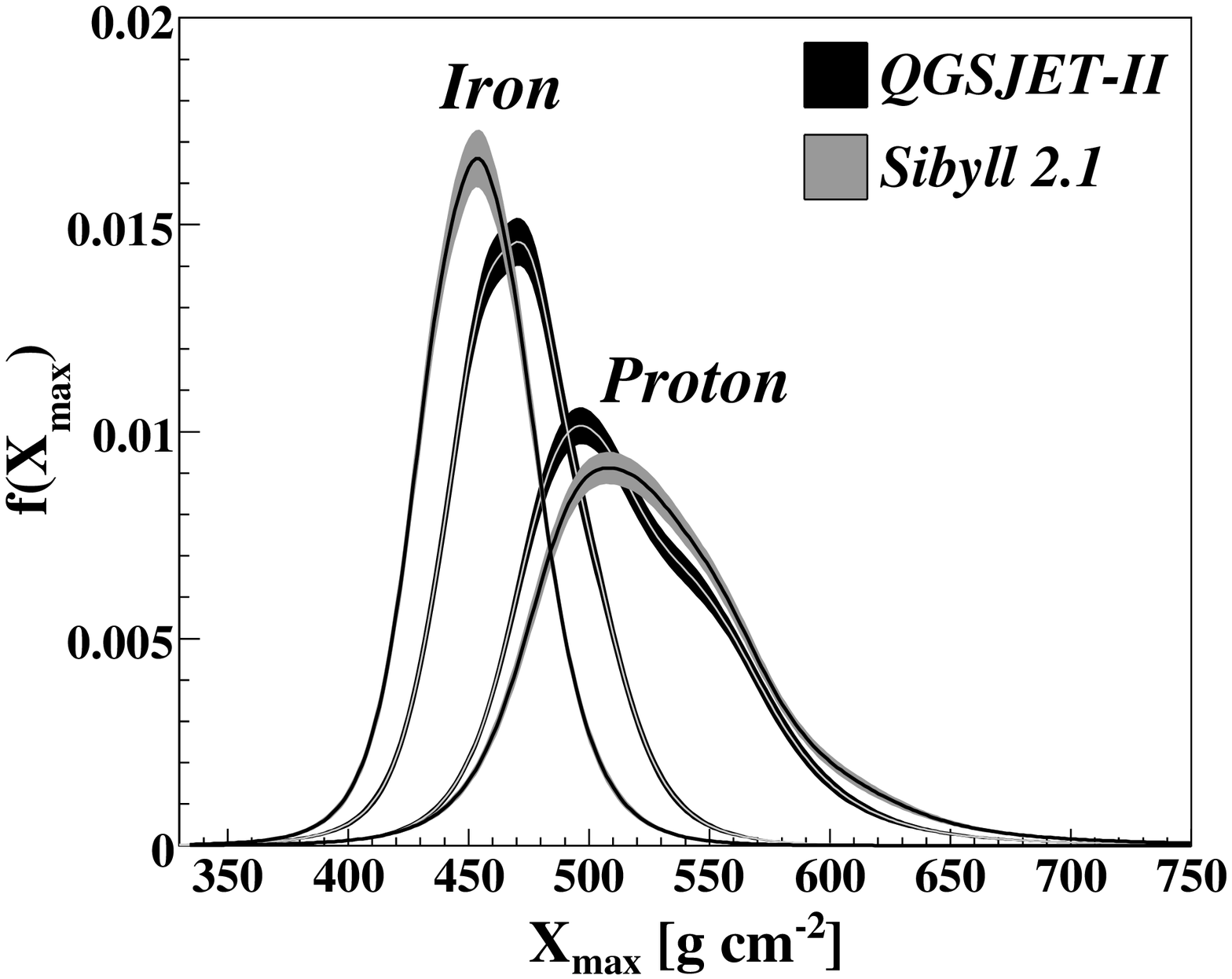}
\caption{Mean value and the one sigma region corresponding to the marginal distribution estimates of $X_{\mathrm{max}}$ 
and $N_{\mu}(600)$ for $\theta=45^\circ$, protons and iron nuclei and QGSJET-II and Sibyll 2.1 as the high energy hadronic 
interaction models.
\label{OneSigma}}
\end{center}
\end{figure}

Fig. \ref{OneSigma} shows the mean value and the one sigma region for the marginal distributions (see Eq. \ref{faa})
of the parameters $X_{\mathrm{max}}$ and $N_{\mu}(600)$ for protons and iron nuclei, $\theta=45^{\circ}$ and QGSJET-II 
and Sibyll 2.1 as the high energy hadronic interaction models. From this figure we see that, on average, QGSJET-II 
produce more muons than Sibyll 2.1 and that the difference between the $\langle X_{\mathrm{max}}\rangle$ of protons 
and iron nuclei is smaller for QGSJET-II.

\subsection{Composition determination}
\label{CompoDet}

The method presented here is statistical. In principle, we want to infer from a data sample the average composition 
of the cosmic rays assuming a mixture of proton and iron nuclei, i.e., $0 \leq c_p \leq 1$. We consider samples of 
$N=100$ and $N=1000$ events which are the number of hybrid and surface detector events in the energy interval considered, 
respectively, expected for the 750 m-array in two years of data taking \cite{Medina:06}. $N=1000$ is also the sample 
size of the hybrid events that fall in the energy interval under consideration for the life time of Auger, $\sim 20$ 
years.

For each pair of mass sensitive parameters, zenith angle, hadronic interaction model and for each value of $c_p$ from 
$0$ to $1$ in steps of $\Delta c_p=0.1$, we sample the average proton and iron distributions to generate 
$1000$ independent samples of $N=100$ events and $300$ independent samples of $N=1000$ events.

For each of these samples, corresponding to a given pair of parameters $(q_1, q_2)$, we calculate $\xi_{q_{1}}$ 
and $\xi_{q_{2}}$ from,
\begin{equation}
\xi_{q_{\nu}}^{kl}(c_{p}) = \frac{1}{N}\ \left[ \sum^{N c_{p}}_{i=1}%
P_{p, \nu}^{kl}( q_{\nu i}^{p} ) + \sum^{N (1-c_{p})}_{i=1}%
P_{p, \nu}^{kl}( q_{\nu i}^{fe} ) \right],
\label{XiCal}
\end{equation}
where $\nu = 1,2$ and
\begin{equation}
P_{p, \nu}^{kl}( q_{\nu} ) = \frac{\hat{f}_{p, \nu}^{k}( q_{\nu} )}{\hat{f}_{p, \nu}^{k}(%
 q_{\nu} )+\hat{f}_{fe, \nu}^{l}( q_{\nu} )}.
\label{Pkl}
\end{equation}
Here $\hat{f}_{p, \nu}^{k}( q_{\nu} )$ is the $k$th marginal distribution function for protons corresponding to the
parameter $q_\nu$ calculated from the $k$th estimate $\hat{f}_{p}^{k}( q_1, q_2 )$. $\hat{f}_{fe, \nu}^{l}( q_{\nu} )$
is the $l$th marginal distribution but for iron nuclei. We take QGSJET-II as the ``true'' high energy hadronic interaction 
model, therefore, we use just the QGSJET-II estimates to calculate $P_{p, \nu}^{kl}( q_{\nu} )$.  

For each sample corresponding to a given pair of mass sensitive parameters, hadronic interaction model, zenith angle and 
proton abundance we obtain $110 \times 110$ values of $\xi_{q_\nu}$ which correspond to all possible combinations of the 
marginal distribution functions of proton and iron nuclei. Finally, for each pair of parameters, hadronic interaction 
model, zenith angle and $c_p$ we calculate the mean values $\langle \xi_{q_{1}} \rangle$, $\langle \xi_{q_{2}} \rangle$, 
the standard deviations $\sigma(\xi_{q_{1}})$, $\sigma(\xi_{q_{2}})$ and the covariance $cov(\xi_{q_{1}},\xi_{q_{2}})$.

\begin{figure}[!bt]
\begin{center}
\includegraphics[width=6.8cm]{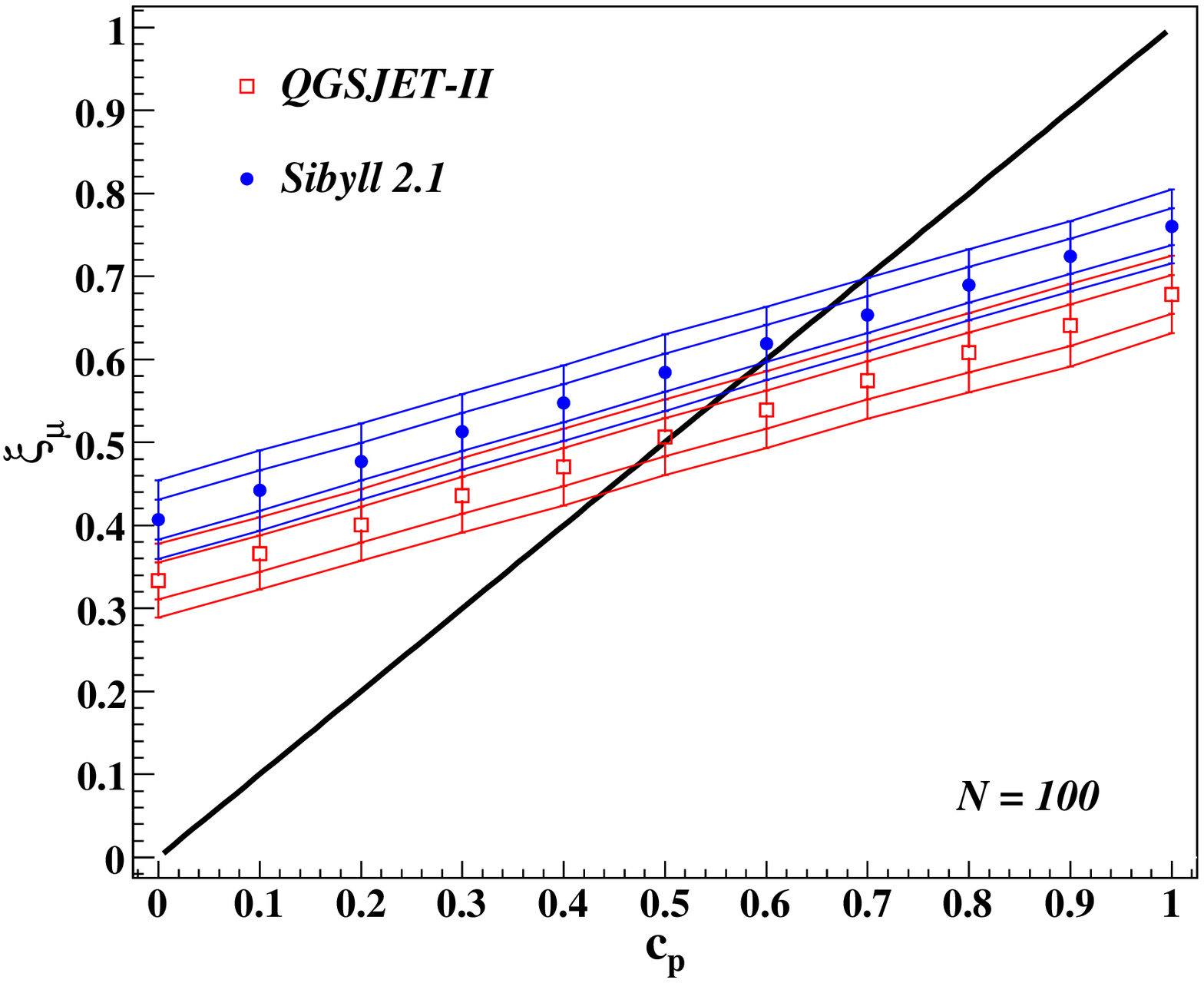}
\includegraphics[width=6.8cm]{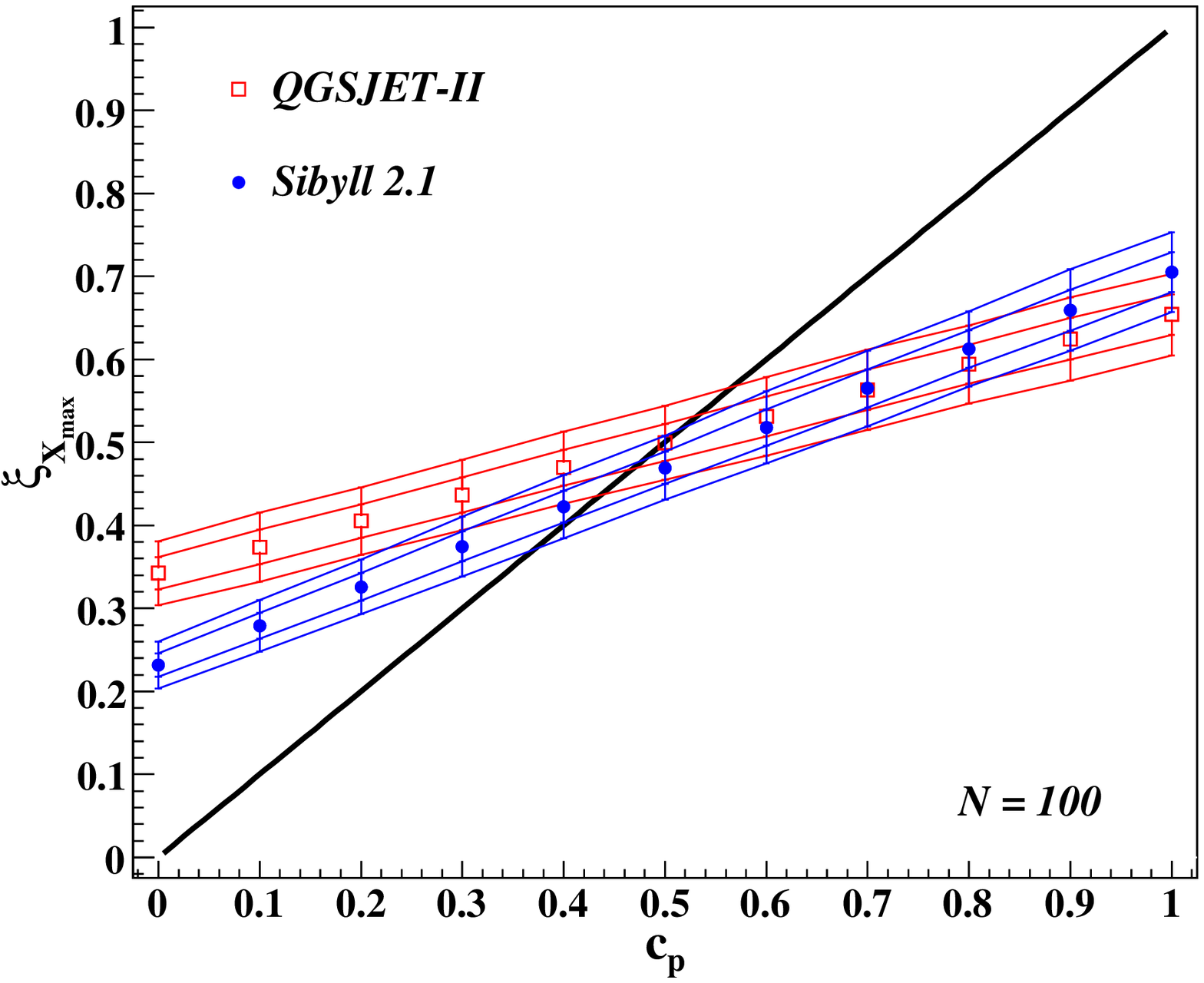}
\includegraphics[width=6.8cm]{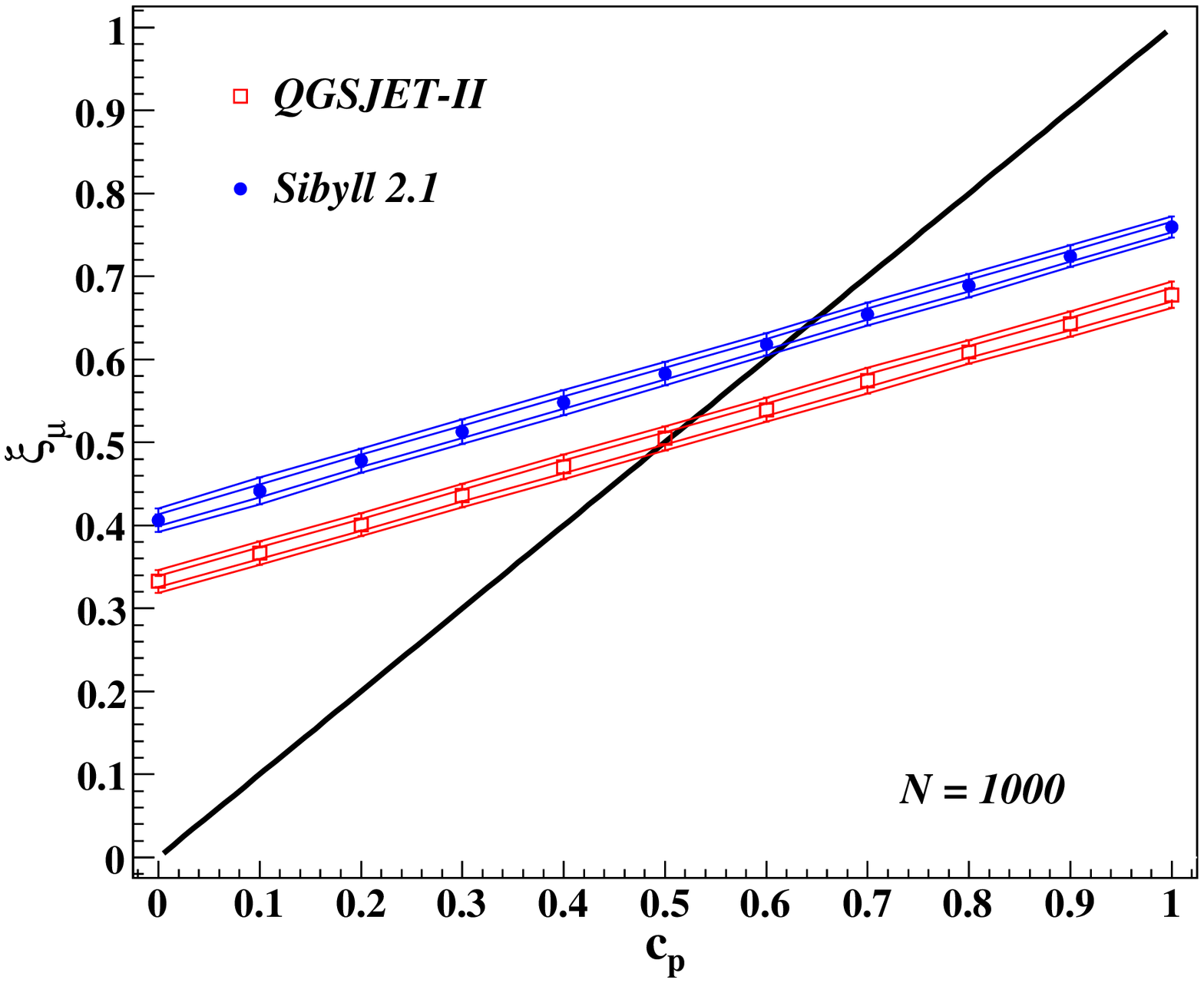}
\includegraphics[width=6.8cm]{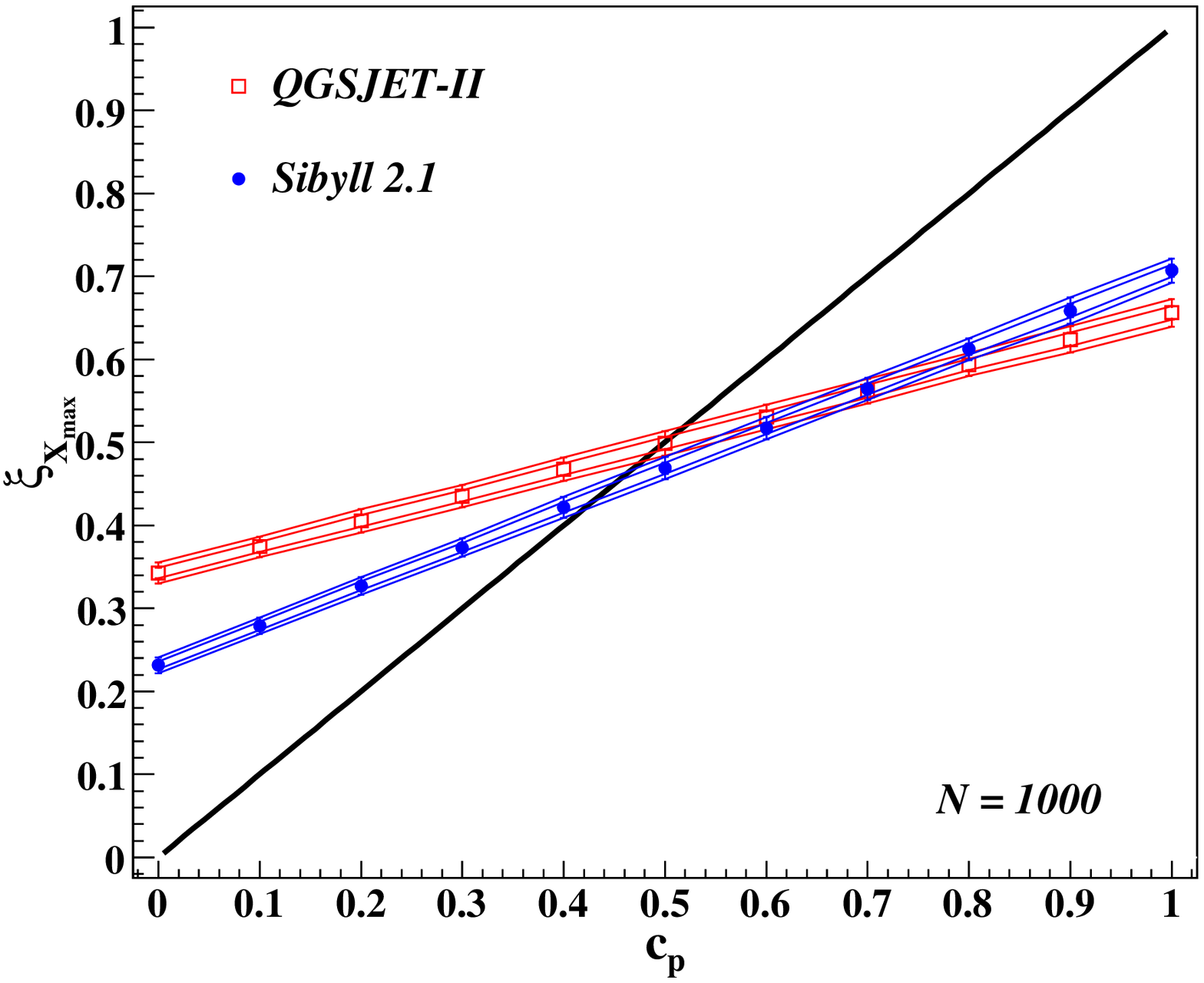}
\caption{Mean value and one and two sigma regions for the parameters $\xi_\mu$ and $\xi_{X_{\mathrm{max}}}$ as a 
function of the proton abundance of the samples for $\theta = 45^\circ$, $N=100$ and $N=1000$ and for samples 
corresponding to QGSJET-II and Sibyll 2.1.
\label{xi45deg}}
\end{center}
\end{figure}

Fig. \ref{xi45deg} shows the parameters $\xi_\mu$ and $\xi_{X_{\mathrm{max}}}$ as a function of the proton abundance of the 
samples for $\theta = 45^\circ$, $N=100$ and $N=1000$ events and samples generated with QGSJET-II and Sibyll 2.1. It also 
shows the one and two sigma regions of $\xi_q$. As expected, $\sigma$ is much smaller for $N=1000$, because, 
as explained in 
section \ref{AbEst}, the variance of $\xi$ is inversely proportional to the sample size.

From Fig. \ref{xi45deg} we see that the mean value of $\xi$ increases linearly with the proton abundance of the
samples, but with slope smaller than one, which corresponds to the ideal case where the iron and proton distributions 
do not overlap. The slopes of $\langle \xi_\mu \rangle$ and $\langle \xi_{X_{\mathrm{max}}} \rangle$ for the QGSJET-II 
samples are comparable, meaning that the discrimination power of $N_{\mu}(600)$ and $X_{\mathrm{max}}$ is similar. This 
is consistent with the results of Ref. \cite{SupaRec:08}, where we showed that the discrimination power of 
$N_{\mu}(600)$ is considerably larger than for $X_{\mathrm{max}}$. But when we take the energy uncertainty
into account, they become comparable since showers' muon content depends almost linearly on primary energy.

For samples generated with Sibyll 2.1 (as mentioned, we take QGSJET-II as the reference model) we see that the behavior 
of $\xi_q$ is quite different, in particular for $c_p=1/2$ the value of $\xi_q$ is no longer $1/2$. Moreover, the mean 
value of $\xi_\mu$, for a given value of $c_p$, is larger for Sibyll 2.1 since, on average, QGSJET-II produces more 
muons than Sibyll 2.1 (see Fig. \ref{OneSigma}). Focusing on $X_{\mathrm{max}}$, we see that for smaller values of 
$c_p$ (larger values of iron abundances), the mean value of $\xi_{X_{\mathrm{max}}}$ is smaller for Sibyll 2.1. This 
is due to the fact that $X_{\mathrm{max}}$ for iron nuclei is, on average, smaller for Sibyll 2.1 
(see Fig. \ref{OneSigma}). For larger values of $c_p$, $\langle \xi_{X_{\mathrm{max}}} \rangle$ for Sibyll 2.1 is of 
order or even larger than the corresponding parameter in QGSJET-II. Again, this happens because the mean value of 
$X_{\mathrm{max}}$ for Sibyll 2.1 corresponding to protons is slightly larger than for QGSJET-II. As a final result, 
the slope of the straight line resulting from Sibyll 2.1 is larger than the slope resulting from QGSJET-II.

We found similar results for $\theta = 30^\circ$. In particular, in this case, the slope of $\xi_{X_{\mathrm{max}}}$ is 
greater than that corresponding to $\xi_\mu$, which means that the discriminating power of $\xi_{X_{\mathrm{max}}}$ is 
greater than that corresponding to $N_{\mu}(600)$. Although the behavior of $\xi_\mu$ and $\xi_{X_{\mathrm{max}}}$ for 
$\theta=30^\circ$ and $\theta=45^\circ$ are qualitatively similar, the dependence on the zenith angle is not negligible.

\begin{figure}[!bt]
\begin{center}
\includegraphics[width=10cm]{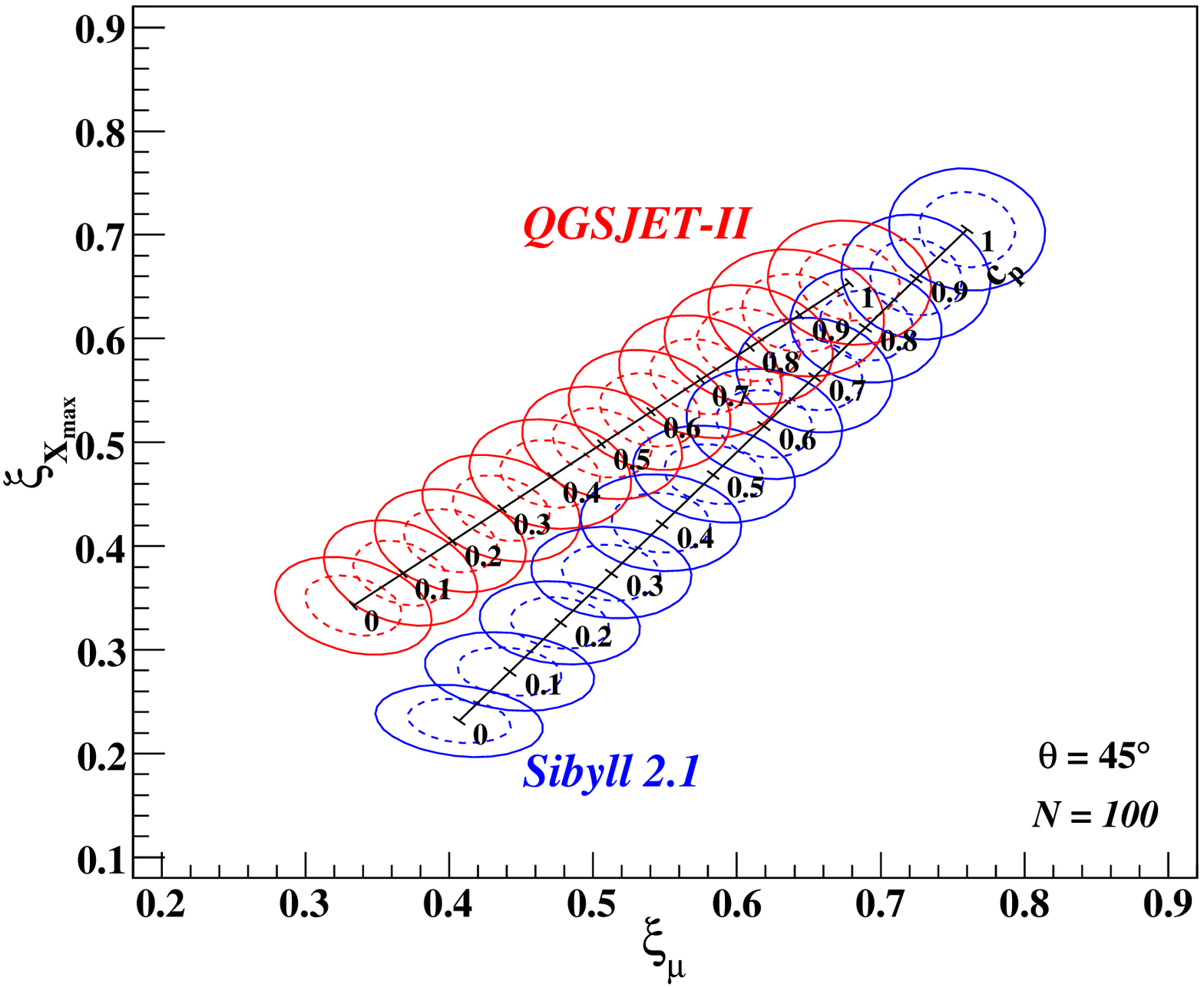}
\includegraphics[width=10cm]{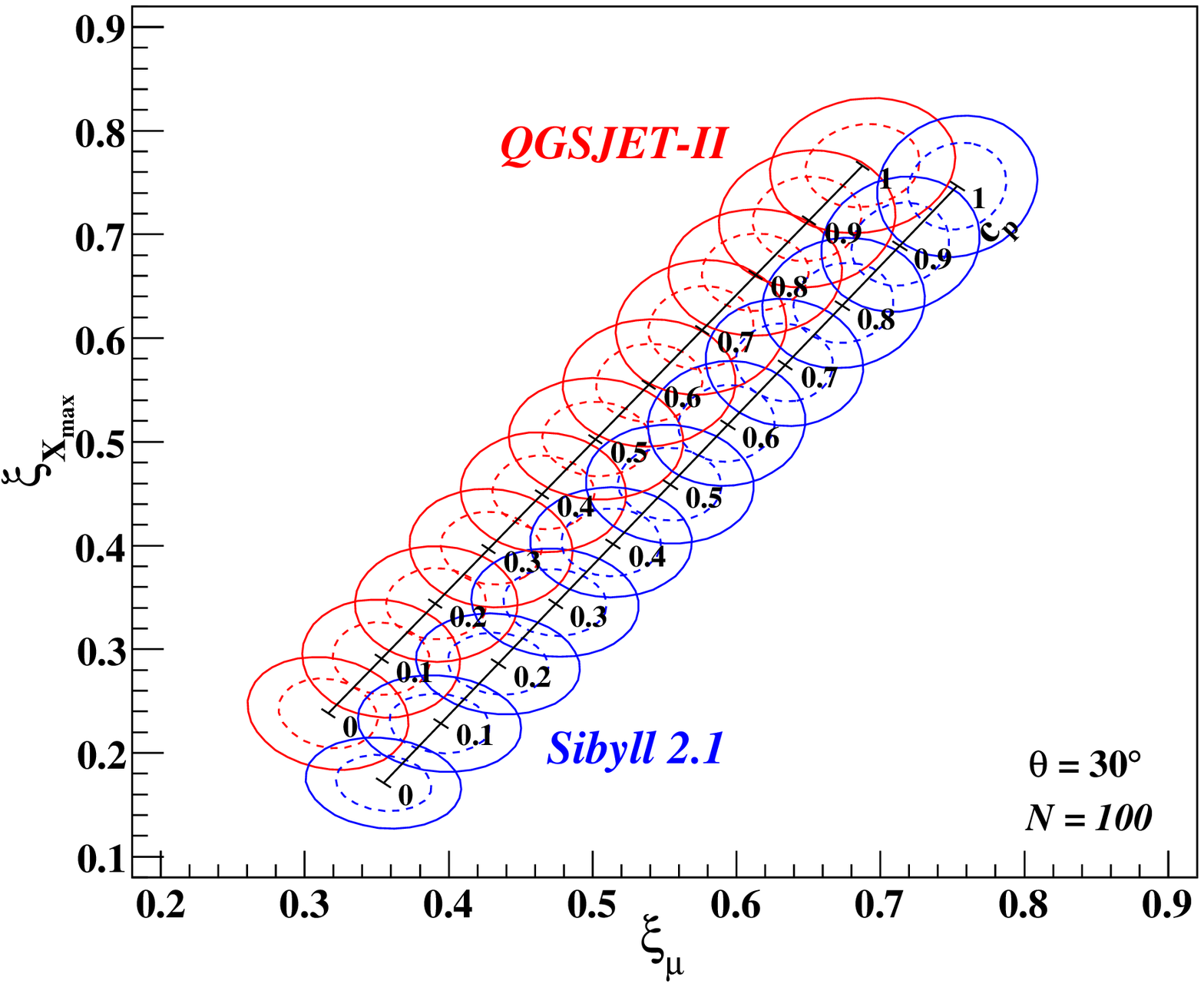}
\caption{Ellipses corresponding to $68\%$ and $95\%$ probability for the Gaussian distributions of the parameters 
$\xi_\mu$ and $\xi_{X_{\mathrm{max}}}$ for $c_{p} \in [0,1]$, $\theta=30^\circ$ and $45^\circ$, $N=100$ events and for 
samples corresponding to QGSJET-II and Sibyll 2.1.
\label{EllXmNm_a}}
\end{center}
\end{figure}

As mentioned in section \ref{AbEst}, the distribution functions of the variables $\xi_{q1}$ and $\xi_{q2}$ 
are Gaussian. The mean value and the covariance matrix depend on the proton abundance of the samples and therefore so 
will the ellipses that enclose regions of a given value of probability. Fig. \ref{EllXmNm_a} shows the ellipses 
corresponding to $68\%$ and $95\%$ probability for the parameters $\xi_\mu$ and $\xi_{X_{\mathrm{max}}}$ for our case 
study: $\theta=30^\circ$ and $45^\circ$, $N=100$ events, for samples built using QGSJET-II and Sibyll 2.1. The evolution 
of the abundance on the $\xi_\mu-\xi_{X_{\mathrm{max}}}$ plane, and the shape and size of the associated ellipses, allow 
for a smooth estimation of the composition in a way that is reasonably independent of the assumed hadronic interaction 
model. Furthermore, given two possible interaction models and an observed data set, a figure like the ones depicted in 
figure \ref{EllXmNm_a} can be used to assess simultaneously the compatibility of these models and the experimental data. 

Having an experimental sample, we can obtain a point on the $\xi_\mu-\xi_{X_{\mathrm{max}}}$ plane by using the density 
estimates obtained from simulations. This point, together with a diagram like the one in Fig. \ref{EllXmNm_a} allow for 
a quick evaluation of the compatibility between the experimental data and the hadronic interaction models under 
consideration. Moreover, if the position of the experimental data point on the diagram is compatible with any of the 
hadronic interaction models, one can also obtain a rough estimation of the composition by simple inspection of the 
nearest ellipses. As an example, let us consider the top panel of Fig. \ref{EllXmNm_a} and an experimental point of 
coordinates $(0.4, 0.24)$. For this particular example, one can immediately tell form the position of the point with 
respect to the curves that the data is compatible with Sibyll 2.1 and that the composition is, approximately, in the 
interval $[0, 0.04]$ at $68\%$ confidence level. If, on the other hand, one considers a point like $(0.56, 0.5)$ it is 
not possible to discriminate between QGSJET-II and Sibyll 2.1; however one can still estimate the composition is in the 
interval $[0.5, 0.6]$ at $68\%$ confidence level. A point like $(0.3, 0.6)$, that is located too far from the curves is 
inconclusive from the point of view of composition or hadronic interaction model, but is a strong indicative of large 
systematic errors in the detector.         
    
\begin{figure}[!bt]
\begin{center}
\includegraphics[width=6.8cm]{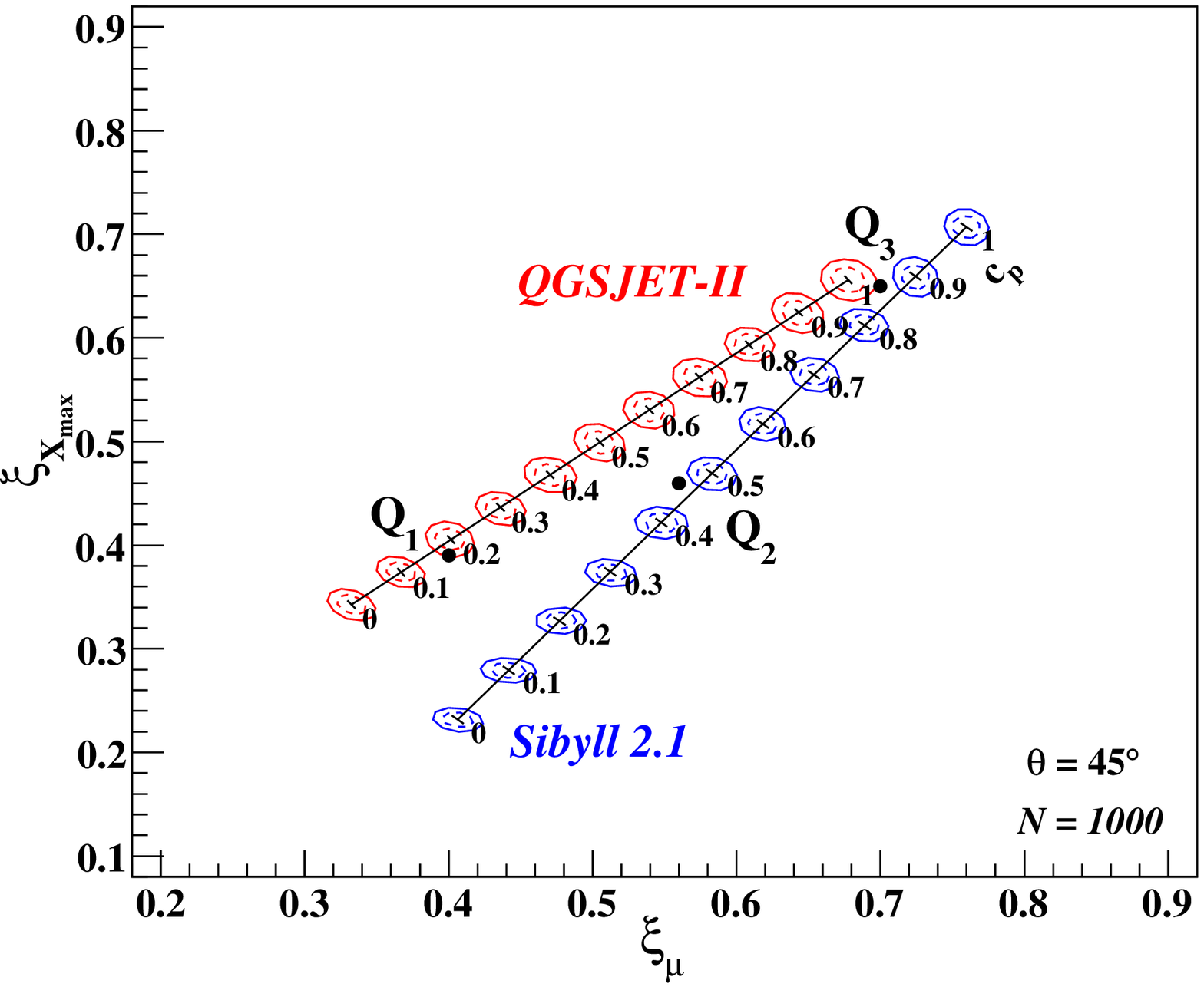}
\includegraphics[width=6.8cm]{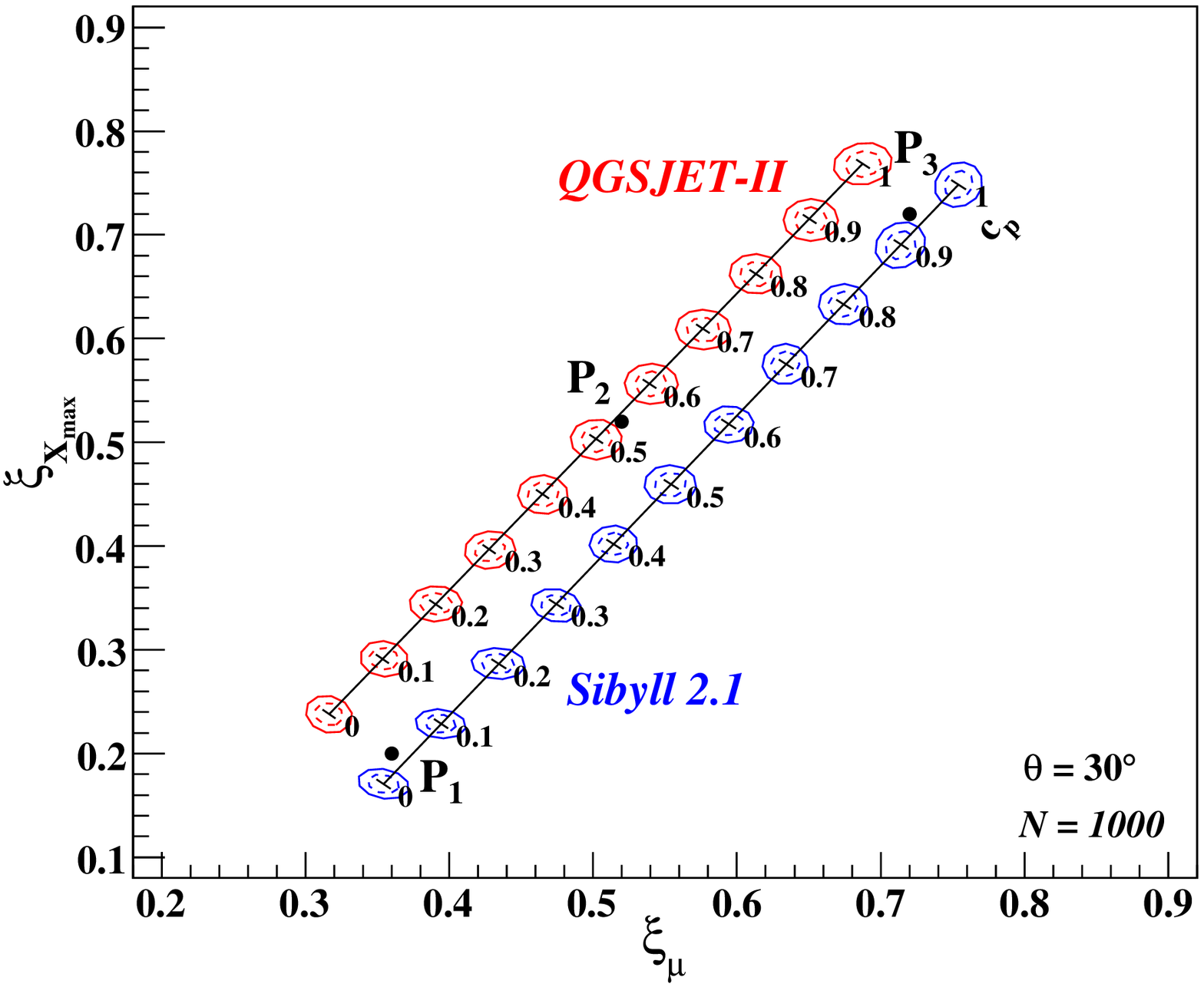}
\caption{Ellipses corresponding to $68\%$ and $95\%$ probability for the Gaussian distributions of the parameters
$\xi_\mu$ and $\xi_{X_{\mathrm{max}}}$ for $c_{p} \in [0,1]$, $\theta=30^\circ$ and $45^\circ$, $N=1000$ events and for
samples corresponding to QGSJET-II and Sibyll 2.1. The points $P_i$ and $Q_i$ with \mbox{$i=1,2,3$} used as examples to 
illustrate the method to infer the proton abundance are also shown.
\label{EllXmNm_b}}
\end{center}
\end{figure}

Fig. \ref{EllXmNm_b} shows the ellipses corresponding to $68\%$ and $95\%$ probability for the parameters $\xi_\mu$ and 
$\xi_{X_{\mathrm{max}}}$: $\theta=30^\circ$ and $45^\circ$, $N=1000$ events, for samples built using QGSJET-II and 
Sibyll 2.1. As expected, the size of the ellipses is smaller than the corresponding to the case of $N=100$ events which 
allow a more stringent test of the compatibility of experimental data with the hadronic interaction models.  

If the experimental point falls close to the region of the ellipses, i.e., the hadronic interaction models are 
compatible with the data, the composition can be estimated in a more formal way. For this purpose we perform a linear 
interpolation of the mean values and the elements of the covariance matrices (see Eqs. (\ref {MVXi},\ref{Cov})), as a 
function of $c_p$ for each zenith angle, sample size and high energy hadronic interaction model considered. We then 
assumed a linear dependence in the $\xi_\mu-\xi_{X_{\mathrm{max}}}$ plane that links the same value of composition, 
$c_{p}$, of the two hadronic models considered (``composition isolines'') and choose $\lambda$ as a variable in the 
$[0, 1]$ interval which takes the values 0 and 1 for QGSJET-II and for Sibyll 2.1, respectively. This approach is 
based on the assumption that a continuous and smooth parametric variation between different hadronic models is possible 
on the $\xi_\mu-\xi_{X_{\mathrm{max}}}$ plane.

In this way we obtain, for each zenith angle and sample size, the functions 
$\vec{\mu}(c_{p},\lambda) = (\langle \xi_{\mu} \rangle (c_{p},\lambda),\langle \xi_{X_{\mathrm{max}}} \rangle (c_{p},\lambda))$
and $\mathbf{V}(c_{p},\lambda)$. The intermediate values of $\lambda$ correspond to hadronic models for which the values 
of the parameters $\xi_\mu$ and $\xi_{X_{\mathrm{max}}}$ fall in between those corresponding to QGSJET-II and Sibyll 2.1. 
Therefore, $\lambda$ parametrizes the composition isolines for hadronic models that would yield points between 
those corresponding to QGSJET-II and Sibyll 2.1.

Let us suppose that for a given experimental sample with zenith angle $\theta$ and number of events $N$ we obtain
$\vec{\xi}^{exp}=(\xi_{\mu}^{exp} , \xi_{X_{\mathrm{max}}}^{exp})$. We can estimate the proton abundance, $\hat{c}_p$, 
by solving the equation $\vec{\mu}(\hat{c}_{p},\lambda) = \vec{\xi}^{exp}$. The regions in the $(c_p, \lambda)$ plane 
compatible with $\vec{\xi}^{exp}$ at a given confidence level are the solutions of the inequality,
\begin{equation}
\left( \vec{\xi}^{exp}-\vec{\mu}(c_{p},\lambda) \right)^{T} \cdot%
\mathbf{V}^{-1}(c_{p},\lambda) \cdot \left( \vec{\xi}^{exp}-\vec{\mu}(c_{p},\lambda)%
\right) \leq R^2(\alpha),
\label{AbCl}
\end{equation}
where $R^2(\alpha) = -2 \ln(1-\alpha)$ and $\alpha$ is the confidence level (for instance,
$\alpha=0.68$, $\alpha=0.95$, etc.).

\begin{figure}[!bt]
\begin{center}
\includegraphics[width=6.8cm]{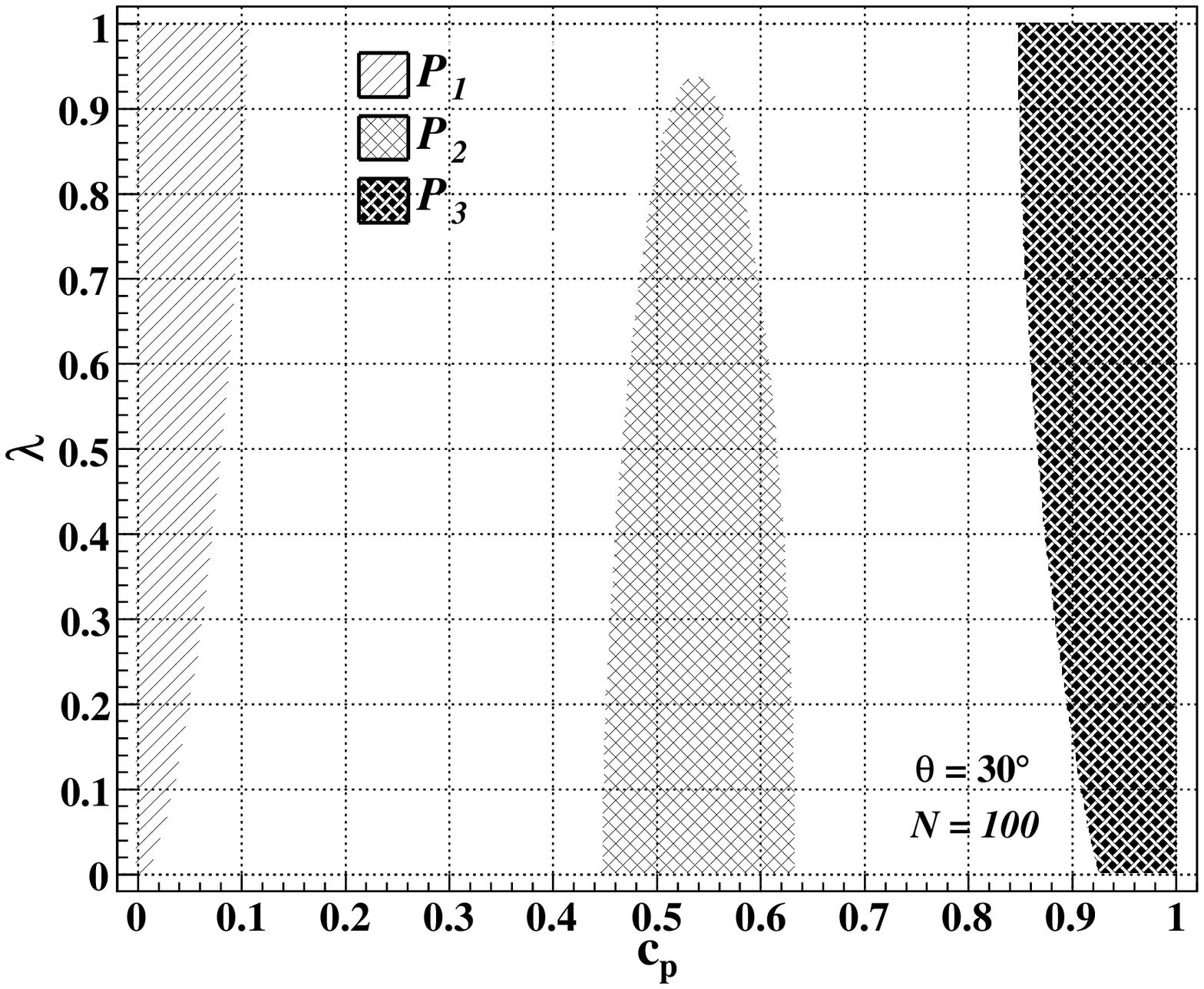}
\includegraphics[width=6.8cm]{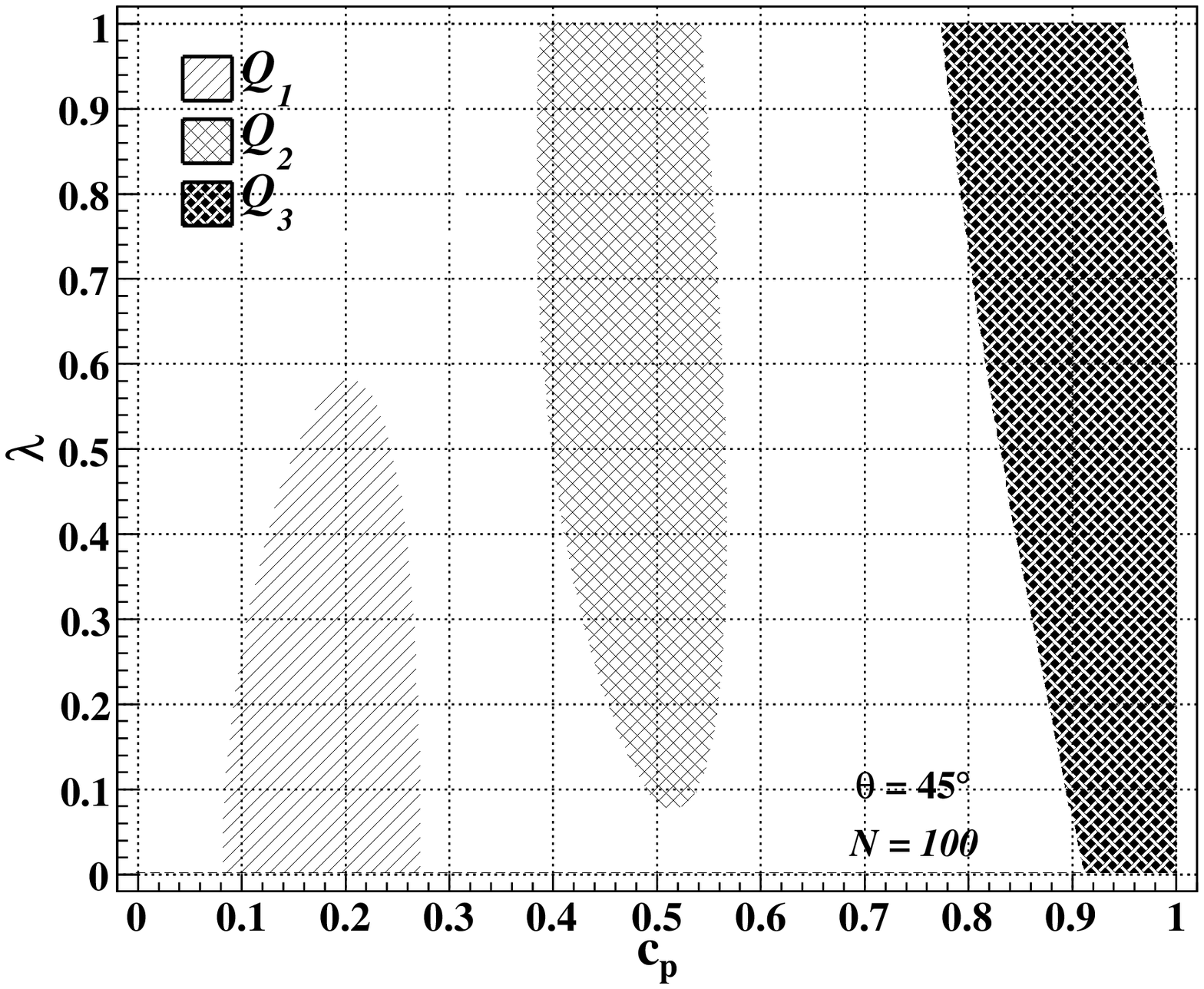}
\includegraphics[width=6.8cm]{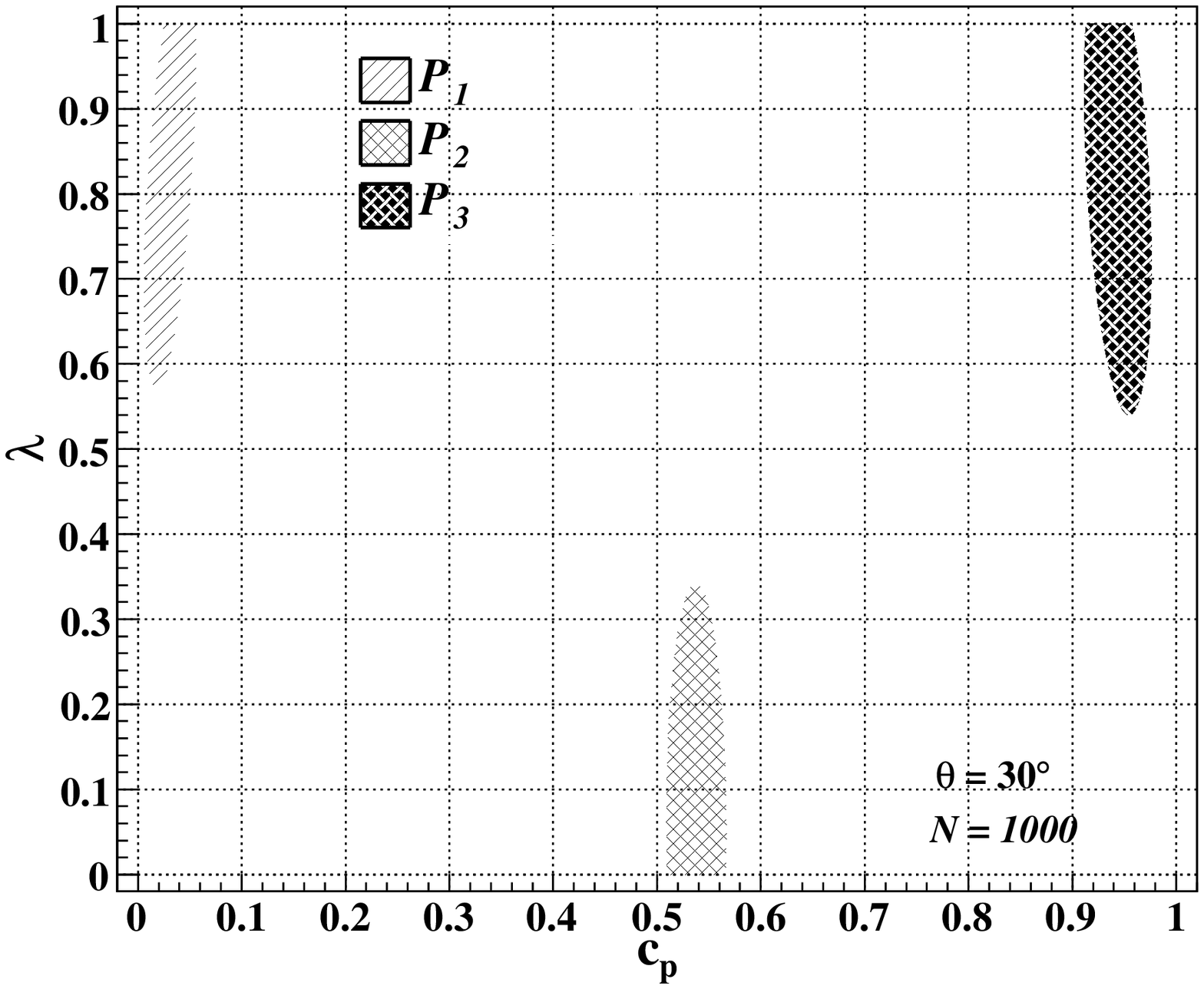}
\includegraphics[width=6.8cm]{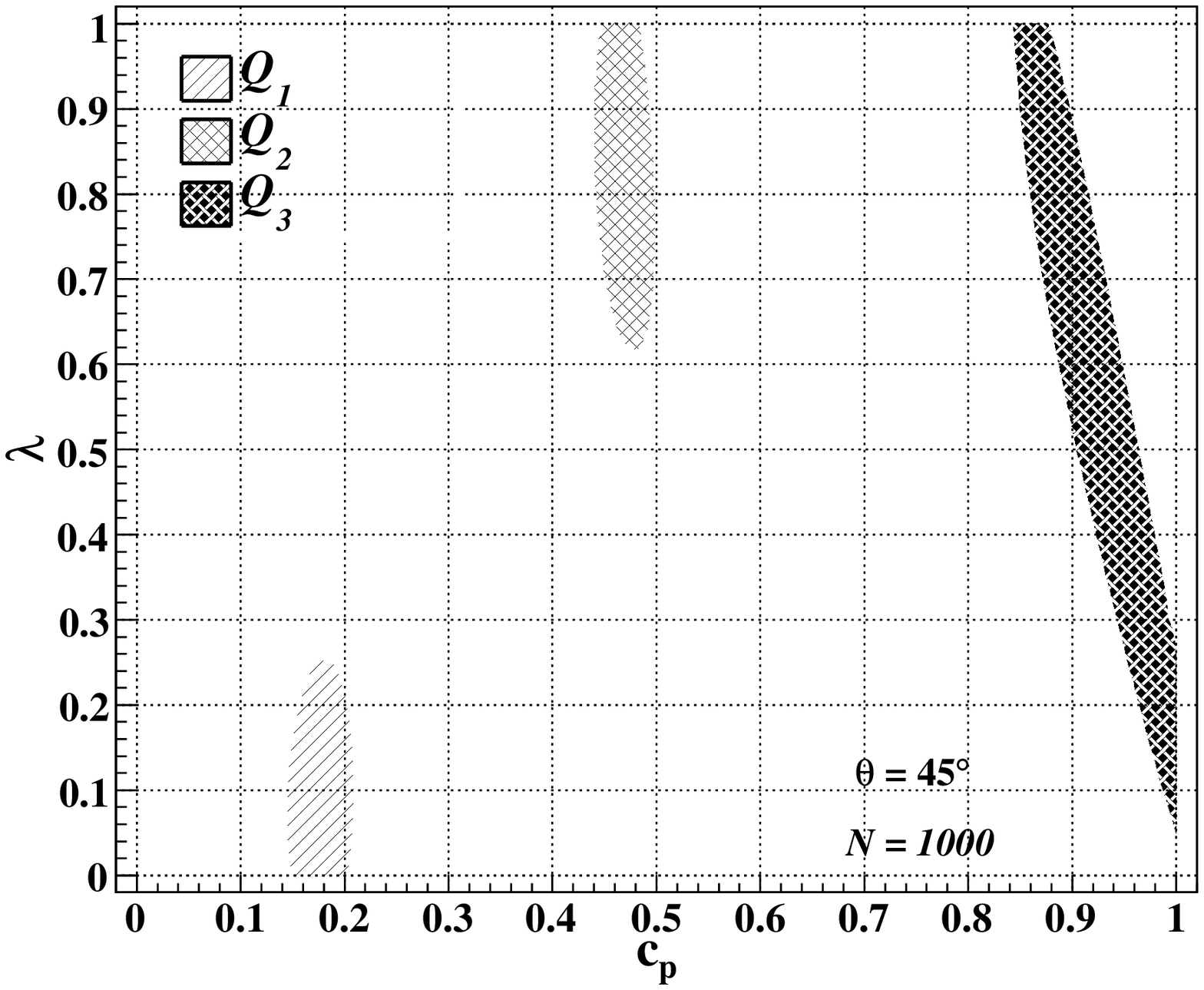}
\caption{Regions in the $c_p-\lambda$ space compatible with the points $P_i$ and $Q_i$ with \mbox{$i=1,2,3$} at
$95\%$ confidence level for samples of $N=100$ and $N=1000$ events.
\label{CL95XmNm}}
\end{center}
\end{figure}

To illustrate the method we calculate the proton abundance of samples of $N=100$ and $N=1000$ events, corresponding 
to the points in the $\xi_\mu-\xi_{X_{\mathrm{max}}}$ space: $P_1 = (0.36 , 0.2)$, $P_2 = (0.52 , 0.52)$ and 
$P_3 = (0.72 , 0.72)$ for $\theta = 30^{\circ}$ and $Q_1 = (0.4 , 0.39)$, $Q_2 = (0.56 , 0.46)$ and $Q_3 = (0.7 , 0.65)$ 
for $\theta = 45^{\circ}$. These points are depicted in the right and left panel of Fig. \ref{EllXmNm_b}, 
respectively. Fig. \ref{CL95XmNm} shows the regions corresponding to $95\%$ confidence level obtained by using Eq. 
(\ref{AbCl}) for $N = 100$ and $N=1000$ sample sizes.

In order to obtain the $c_p$ intervals compatible, at $95\%$ confidence level, with the considered points, we have to 
project the regions of Fig. \ref{CL95XmNm} onto the $x$ axis, corresponding to proton abundance. Table \ref{CompoInf} 
shows the inferred composition and its uncertainty at $95\%$ confidence level. Note that until now we just consider the 
composition of a given sample of cosmic rays, not the one corresponding to the universe (sample of infinite number of 
events), to see how to obtain the proton abundance of the cosmic rays from the composition of a sample see Appendix 
\ref{CRCompDet}.
\begin{table}[h]
\begin{center}
\caption{Inferred proton abundance of the samples and its uncertainty at $95\%$ confidence level for the points
$P_i$ and $Q_i$ with \mbox{$i=1,2,3$} for samples of $N=100$ and $N=1000$ events.}
\label{CompoInf}
\begin{tabular}{c c c} \hline
Points &  $c_{p}^{inf}$ for $N = 100$ &  $c_{p}^{inf}$ for $N = 1000$  \\  \hline
$P_1$  & $0.031^{+ 0.076}_{-0.031}$ &  $0.031 \pm 0.025$ \\ 
$P_2$  & $0.537^{+ 0.096}_{-0.090}$ & $0.537 \pm 0.029$ \\  
$P_3$  & $0.945^{+ 0.055}_{-0.097}$ & $0.944 \pm 0.033$ \\  
$Q_1$  & $0.177 \pm 0.095$ &  $0.176 \pm 0.032$ \\  
$Q_2$  & $0.468^{+ 0.099}_{-0.083}$ &  $0.469 \pm 0.029$ \\  
$Q_3$  & $0.90^{+ 0.10}_{-0.13}$ &  $0.906^{+ 0.094}_{-0.062}$ \\  \hline
\end{tabular}
\end{center}
\end{table}

For $\theta=30^\circ$ ($P$-points) we see that the uncertainty in the determination of the composition varies very slowly 
with the proton abundance of the samples, in particular, the error increases for values of $c_p$ closer to one. For 
$\theta=45^\circ$ ($Q$-Points) the uncertainty varies faster, taking larger values than for $\theta=30^\circ$ in the 
$c_p=1$ region due to the superposition of the ellipses.

\begin{figure}[!bt]
\begin{center}
\includegraphics[width=6.8cm]{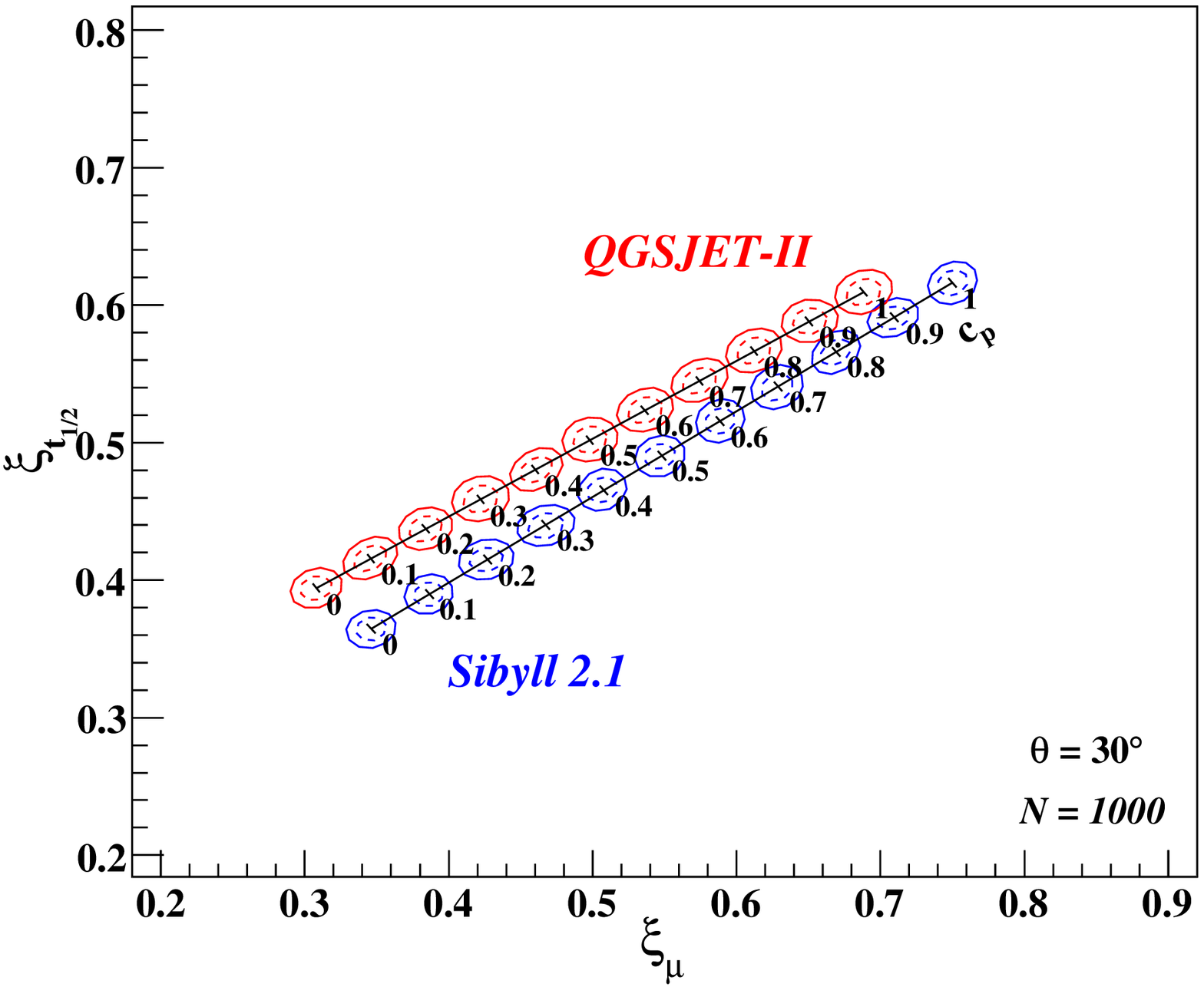}
\includegraphics[width=6.8cm]{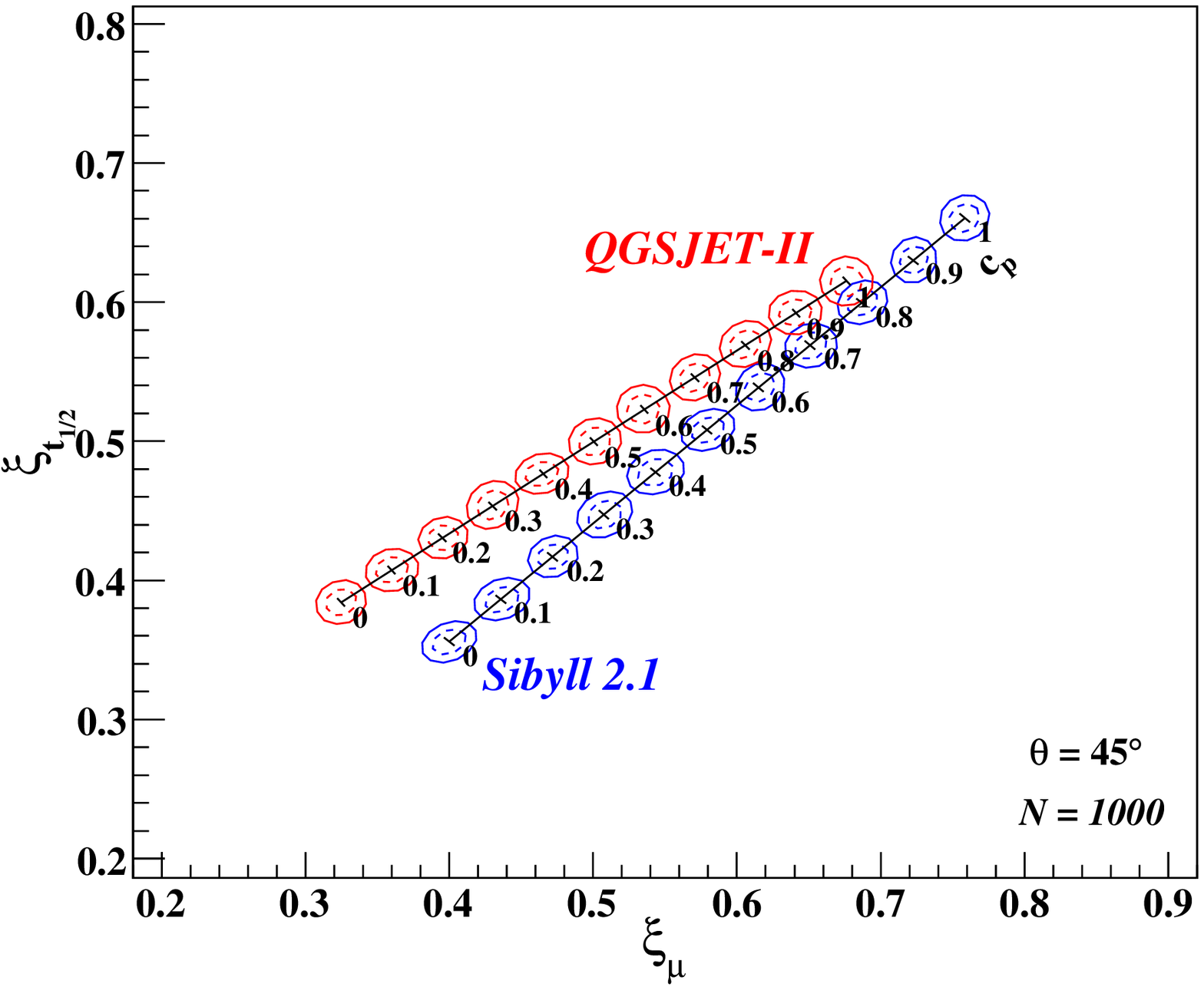}
\includegraphics[width=6.8cm]{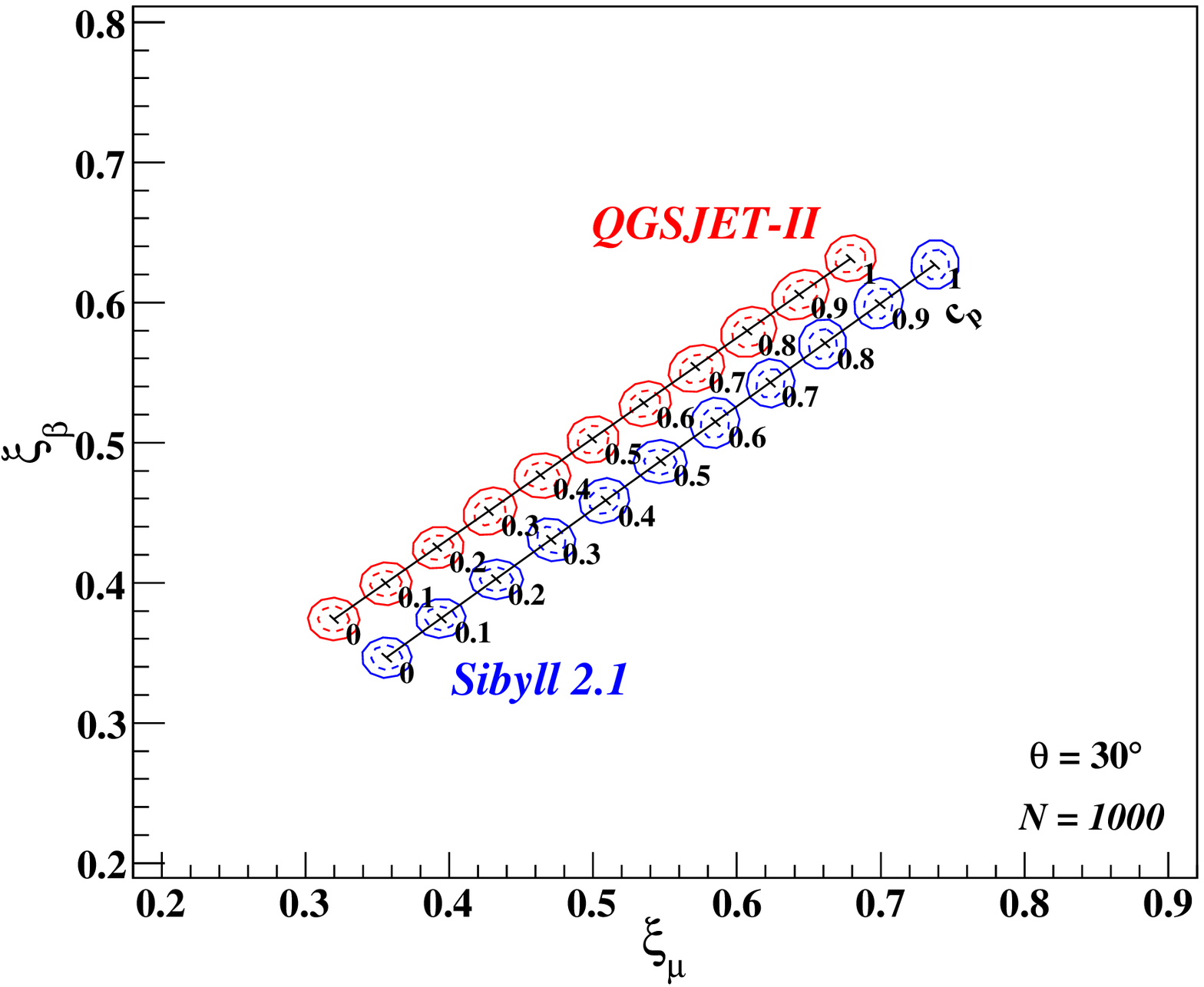}
\includegraphics[width=6.8cm]{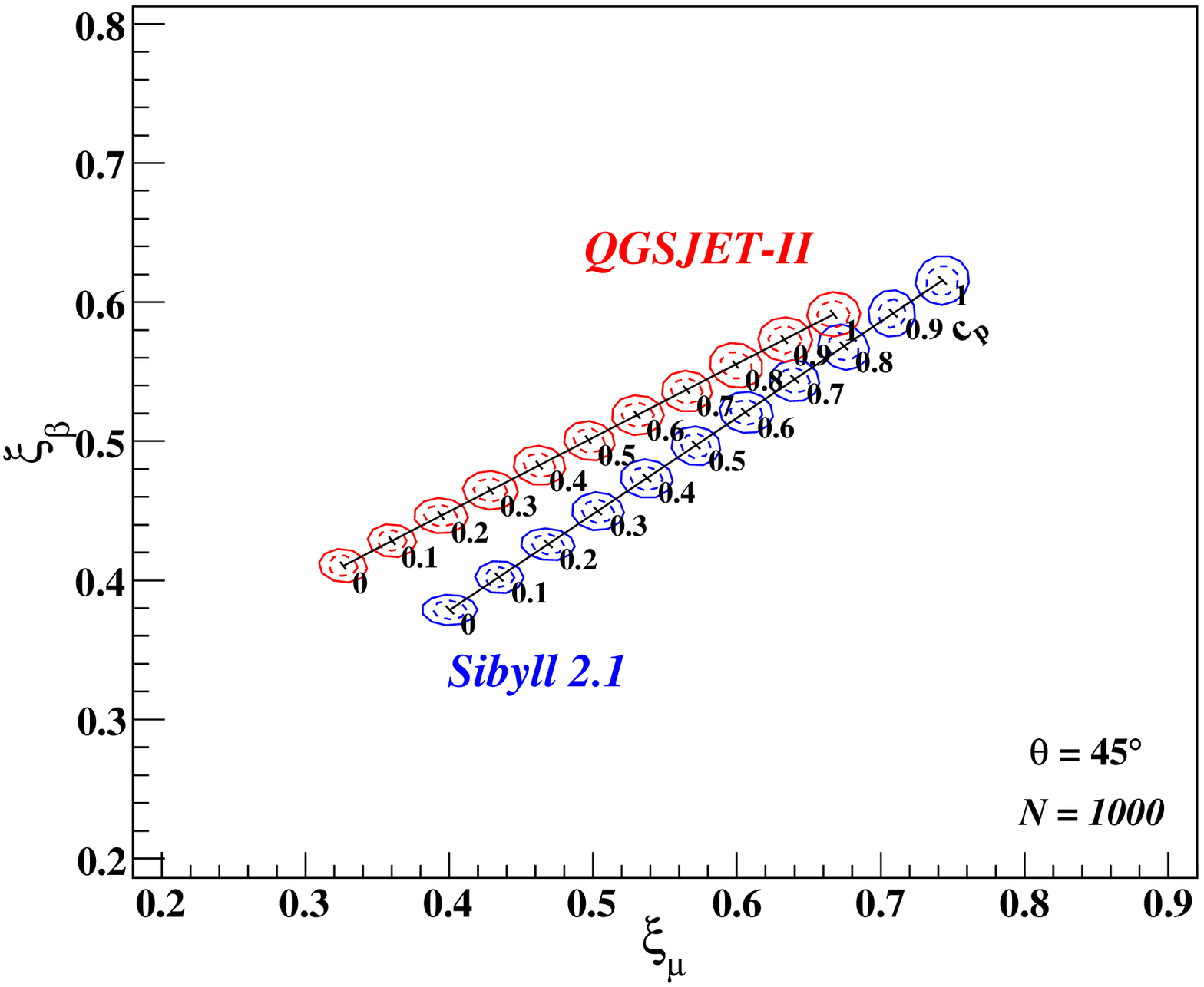}
\includegraphics[width=6.8cm]{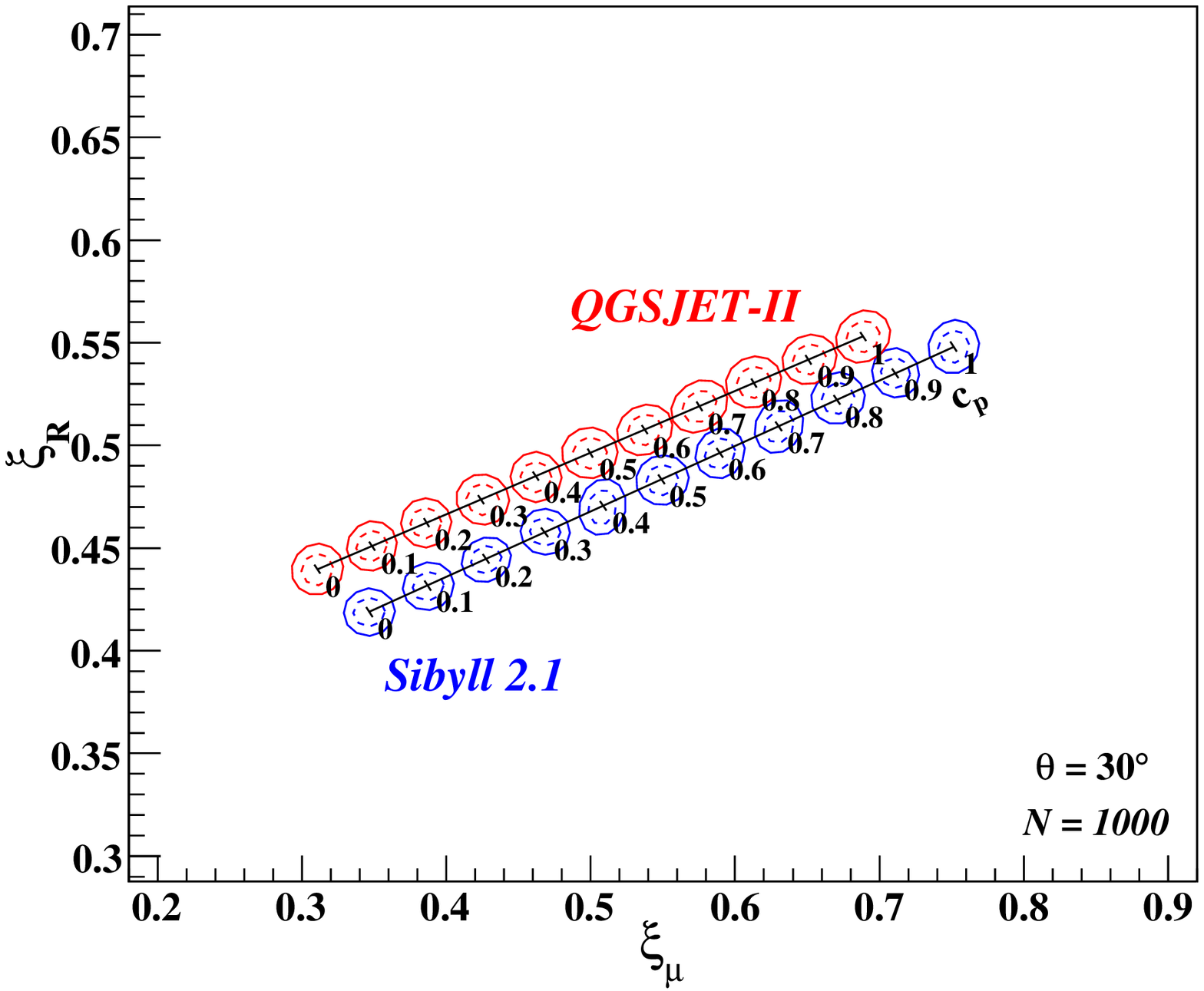}
\includegraphics[width=6.8cm]{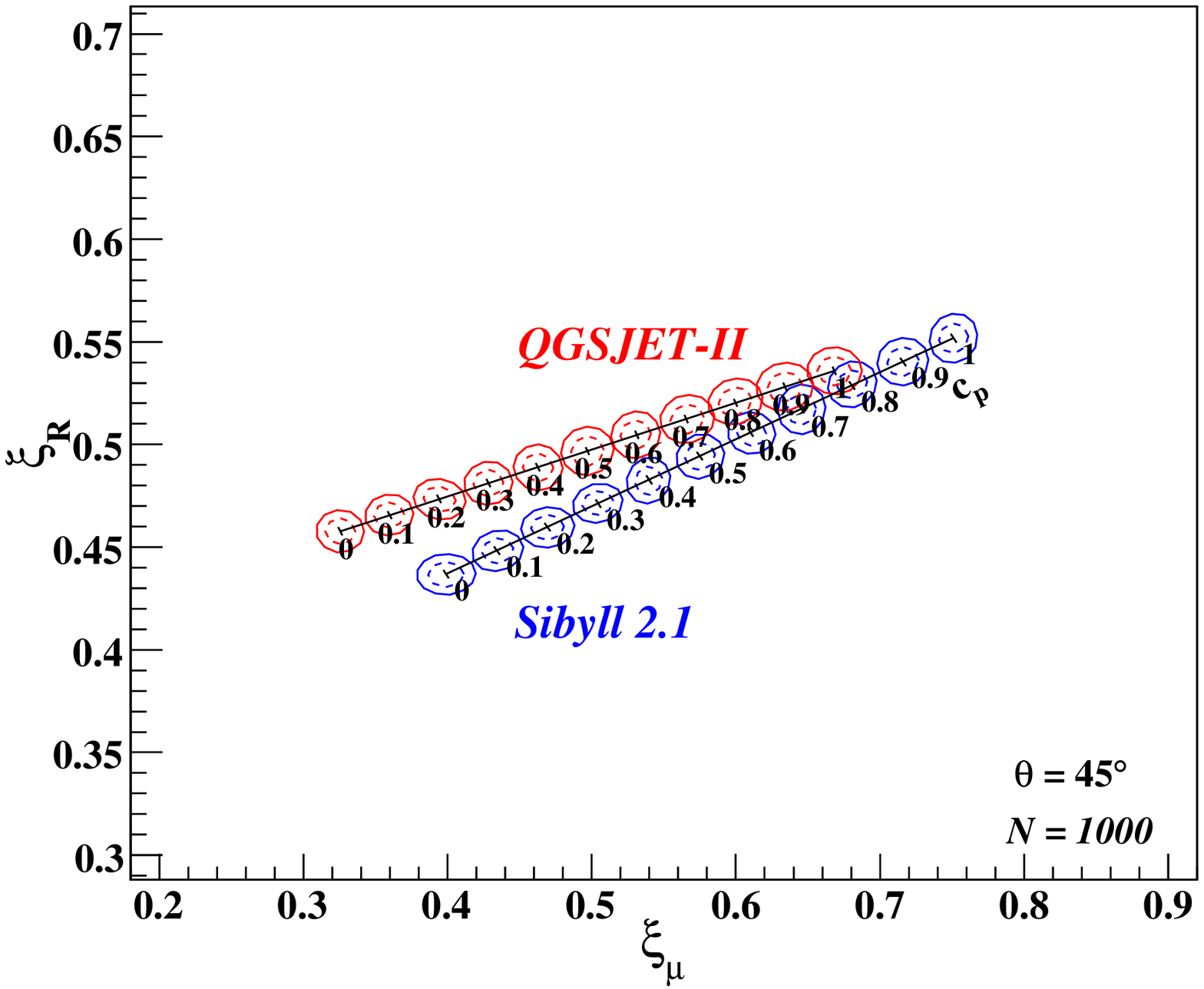}
\caption{Ellipses Corresponding to $68\%$ and $95\%$ of probability for the pair of parameters 
$(\xi_{\mu},\xi_{t_{1/2}})$, $(\xi_{\mu},\xi_{\beta})$ and $(\xi_{\mu},\xi_{R})$, $c_{p} \in [0,1]$, 
$\theta = 30^{\circ}$ and $\theta = 45^{\circ}$, $N = 1000$ and for samples generated with the hadronic 
interaction models QGSJET-II and Sibyll 2.1.}
\label{EllPar30deg}
\end{center}
\end{figure}

The Auger Observatory measures other composition-sensitive parameters apart from $X_{\mathrm{max}}$ and $N_\mu(600)$ by 
means of the water Cherenkov detectors. Although these parameters strongly depend on $X_{\mathrm{max}}$ and $N_\mu(600)$, 
Cherenkov detectors (and muon counters) have $100\%$ duty cycle, as opposed to fluorescence telescopes, which have only 
$10\%$ duty cycle. This fact highlights the statistical value of surface parameters. For this reason we also studied the 
applicability of our method to the combination of $N_\mu(600)$ with other parameters obtained from the Cherenkov detectors: 
$t_{1/2}$, $\beta$ and $R$ (see subsection \ref{DenEst}). Fig. \ref{EllPar30deg} shows the ellipses corresponding to $68\%$ 
and $95\%$ probability for the combination of $\xi_{\mu}$ with $\xi_{t_{1/2}}, \xi_{\beta}$ and $\xi_{R}$. We consider 
samples of $1000$ events which is the number of events expected in two years of data taking for the 750 m-array of AMIGA.

Fig. \ref{EllPar30deg} shows that, also in these cases, the position of the ellipses allows 
us to test the high 
energy hadronic models as well as to obtain the proton abundance of a sample, when the experimental data falls in 
the proximity of the ellipses. Again, the composition uncertainty increases in the region close to $c_p=1$ especially 
for $\theta = 45^\circ$.

A note must be made regarding the potential effects of systematic errors on our technique. Reconstruction biases are 
rather well known and, very likely, included in a fairly acceptable way in the reconstruction packages used. Furthermore, 
all the distributions use here are based on reconstructed simulated data. The energy bias, on the other hand, is a 
fundamental problem for high energy cosmic ray shower measurement at present and it is, certainly, a potential problem 
for any technique that attempts to infer cosmic ray composition from extensive showers at high energies. Our technique 
is not an exception in this respect, specially because we use muon number as one of our main parameters which is 
particularly sensitive to energy. Nevertheless, there are experimental and theoretical facts that can attenuate the 
problem, or even give it a potentially interesting twist: (i) the bias in energy is bounded to around 30\% or, at most, 
50\% and (ii) the sign of the bias is known, i.e., Monte Carlo simulations overestimate the energy and not the other way 
around, (iii) the difference in muon content between hadronic interaction models in the ankle region is about 5\%, (iv) 
$X_{\mathrm{max}}$ has a logarithmic dependence on energy. This means that, in the $\xi_{X_{\mathrm{max}}}-\xi_{\mu}$ space, 
curves move in a predictable direction due to the bias and by an amount that is bounded but larger than the expected one 
due to uncertainties in the hadronic interaction model. Furthermore, the displacement is mostly parallel to the $\xi_{\mu}$ 
axis, without changing appreciably the distance between points of equal composition along curves of constant hadronic 
interaction model. This implies that, even in the presence of an energy bias, the position of an experimental point on 
this plane still carries significant physical information. In fact, while one would lose the capacity of discriminating 
between hadronic interaction models for large systematic biases in energy, one could gain the capacity of experimentally 
constraining the bias with an uncertainty of the order of 5\%. 

In any case, the existence of systematic errors is an important and by no means trivial problem, which deserves further 
studies and the search for more appropriate strategies \cite{Supanitsky:09}, like the modification or refinement of the 
parameters actually used in the context of the technique proposed here. Therefore, one alternative line of action, might 
be, for example, the redefinition of a muon content parameter in the way proposed by Hillas \cite{Hillas:71}.

\section{Conclusions}

Detailed composition studies at the energies of the second knee and ankle of the cosmic ray spectrum will be crucial to 
weight different astrophysical models of the Galactic-extragalactic cosmic ray flux transition. Experiments like Auger and 
Telescope Array in the near future will be instrumental on the later. 

In this paper we present a new statistical method to perform composition studies in a two dimensional space. The method was 
designed having in mind the enhancements AMIGA and HEAT presently under construction by the Pierre Auger Observatory, but its 
applicability extends to other detectors, like TALE, which also have hybrid capability.

A main advantage of the method is that it minimizes the effects of the present uncertainty associated with the hadronic 
interaction models, used to simulate cosmic ray showers, on the inferred composition. Furthermore, in the case in which 
systematic errors in energy are known or smaller than 5\%, the method allows, besides the determination of the composition, 
an independent verification of the compatibility between real shower data and hadronic interaction models. In the case of 
larger systematic errors (e.g., $\sim 30$\%), the technique still allows to make reliable composition estimation and, 
additionally, to set an upper limit on the size of the systematic error in energy.

The novelty of the Auger enhancements is the combination, at energies between $\sim 10^{17}$ and $\sim 10^{18.5}$ eV, of 
hybrid measurement with additional simultaneous, and independent muon number information. We exploit this wealth of data by 
working on a two-dimensional space $\xi_{1}$-$\xi_{2}$ which encodes, e.g., $X_{max}$ and $N_{\mu}$ information. This space 
has the property of clearly separating composition and hadronic interaction model dependencies. Confidence levels are 
calculated in the form ellipses that enclose regions of $68\%$ and $95\%$ probability. It is also shown the way in which  
the $\vec{\xi}$ distribution functions decrease as the size of the sample increases in a way compatible with the exposure 
time scale appropriate for AMIGA-HEAT. We show that, for the case in which the data is not dominated by systematic errors 
in energy, the constraints imposed on the hadronic models become stronger as the exposure grows while the composition error 
diminishes to an unprecedented accuracy for astrophysical applications over the lifetime of the experiment. Therefore, 
besides being able to constrain the high energy hadronic models, it should be possible to determine the composition as a 
function of energy in the ankle region with errors varying from $\sim 20$\%  to $\sim 5$\%, at the 95\% confidence level, 
as the data taking progresses from 2 to 20 yr.

\section{Acknowledgments}

The authors have greatly benefited from discussions with several colleagues
from the Pierre Auger Collaboration, of which they are members. We also want
to acknowledge Greg Snow for carefully reading the manuscript and for his valuable 
comments. GMT acknowledges the support of DGAPA-UNAM through grant IN115707.

\appendix

\section{Optimum energy bin}
\label{EnergyBin}

The primary energy is obtained by fitting a lateral distribution function to the total signal in 
each surface station of the water Cherenkov array. This allows us to interpolate the shower signal 
at a fixed distance from the core which, in turn, is used as an energy estimator. This reference distance 
is such that the shower fluctuations go through a minimum in its vicinity, and its exact value depends on 
the geometry of the array; for Auger, the reference distance is $1000$ m for the $1500$ m baseline spacing 
and $600$ m for the AMIGA infill of $750$ m spacing.
 
The signal at the reference distance is calibrated with the telescopes via hybrid events. The corresponding 
energy uncertainty for the 1500 m-array of Auger is $\sim 20\%$ \cite{GZKAuger}. Guided by this experimental 
result, we assume in this work a $25\%$ Gaussian energy uncertainty.

In order to study the composition at energies of order $E_0 = 1$ EeV, we first have to determine 
the range of reconstructed energies, centered at $E_{r0}$, with the form 
$\Pi_{r} = [(1-\delta) E_{r0}, (1+\delta) E_{r0}]$ ($\delta = 0.25$ for $25\%$ of energy uncertainty) such 
that the fraction of events in the interval $\Pi_{0} = [(1-\delta) E_{0}, (1+\delta) E_{0}]$ is maximum. The 
intervals $\Pi_r$ and $\Pi_0$ are different because of the spectrum. The contamination of events of real energies
smaller than $1$ EeV which are outside $\Pi_0$ is greater than the one corresponding to energies, also outside 
$\Pi_0$, but above $1$ EeV.

We assume that the number of cosmic ray showers with real energies between $E$ and $E+dE$ is well
represented by the following power law spectrum,
\begin{equation}
\frac{dN}{dE}(E) = N_{0}\ (\gamma -1) \frac{E_{1}^{\gamma -1} E_{2}^{\gamma -1}}{%
E_{2}^{\gamma -1}-E_{1}^{\gamma -1}}\ E^{-\gamma},
\label{Spec}
\end{equation}
where, for every set of simulated events we have $E_{1} = 0.6$ EeV, $E_{2} = 2$ EeV, 
$N_{0} \cong 6640$ (depending on the reconstruction efficiency) and $\gamma = 2.7$ (see subsection \ref{Sim}).

The number of events whose real energy belongs to $\Pi_0$ such that the reconstructed energy falls in $\Pi_r$ is
given by,
\begin{equation}
f_{A}(E_{r0}) = \int_{(1-\delta) E_{r0}}^{(1+\delta) E_{r0}} dE%
 \int_{(1-\delta) E_{0}}^{(1+\delta) E_{0}} dE' \ \frac{dN}{dE'}(E') \ G(E,E'),
\label{fa}
\end{equation}
where $G(E,E') = \exp[-(E-E')^2/(2 \delta^2 E'^2)]/(\sqrt{2 \pi} \ \delta \ E')$.

On the other hand, the number of events with real energy outside $\Pi_0$ whose reconstructed energy falls
in $\Pi_r$ is given by,
\begin{eqnarray}
f_{B}(E_{r0}) &=& \int_{(1-\delta) E_{r0}}^{(1+\delta) E_{r0}} dE \left[%
\int_{E_{1}}^{(1-\delta) E_{0}} dE' \ \frac{dN}{dE'}(E') \ G(E,E')+\right. \nonumber \\
&&\left. \int_{(1+\delta) E_{0}}^{E_{2}} dE' \ \frac{dN}{dE'}(E') \ G(E,E')\right].
\label{fb}
\end{eqnarray}

\begin{figure}[!bt]
\begin{center}
\includegraphics[width=13cm]{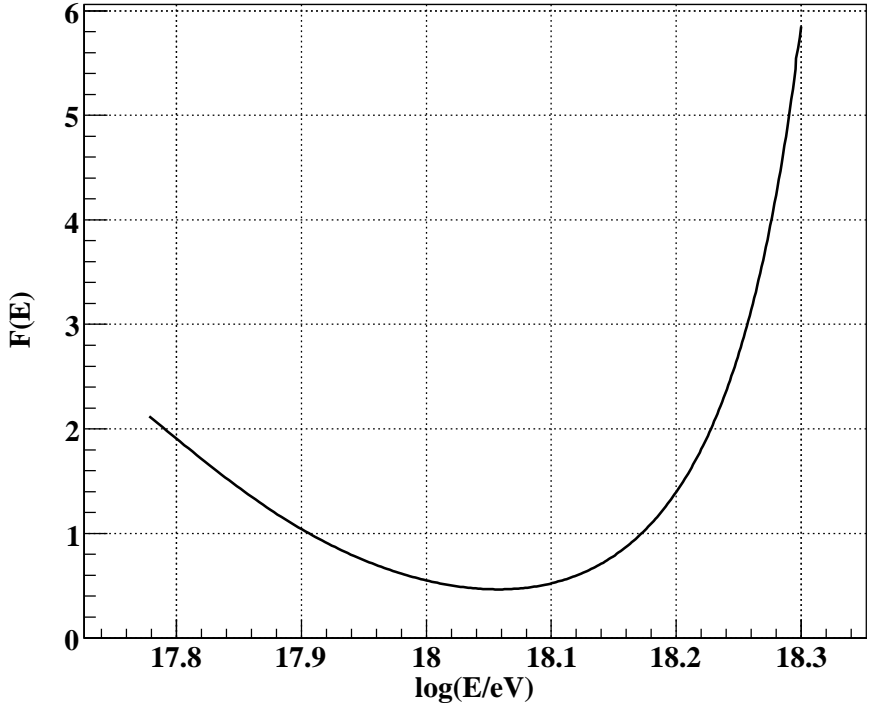}
\caption{Ratio of the number of events whose real energy does not belong to $\Pi_0$ but the reconstructed 
energy falls in $\Pi_r$ and the number of events whose real energy belongs to $\Pi_0$ but the reconstructed 
energy falls in $\Pi_r$, $F(E_{r0})$. The minimum is located at $E_{r0} \cong 1.14$ EeV. \label{BinCenter}}
\end{center}
\end{figure}

Therefore, the value of $E_{r0}$ for which the fraction of events belonging to $\Pi_0$ that fall 
in $\Pi_r$ is maximum is obtained by minimizing the function $F(E_{r0})=f_{B}(E_{r0})/f_{A}(E_{r0})$. 
Fig. \ref{BinCenter} shows $F(E_{r0})$ for $\delta=0.25$ from which we see that it has a minimum but 
at an energy greater than $1$ EeV. The minimum is located at $E_{r0} \cong 1.14$ EeV, 
then, $\Pi_r =[0.86 ,\ 1.43]$ EeV.

\section{Composition of cosmic rays from the composition of a sample}
\label{CRCompDet}

Given a sample of $N$ events, assuming a mixture of protons and iron nuclei, the number of protons in the sample
follows a binomial distribution,
\begin{equation}
P(n_{p};N, \mathcal{C}_{p}) = {N \choose n_{p}} \ \mathcal{C}_{p}^{n_{p}} \ (1-\mathcal{C}_{p})^{N-n_{p}},
\label{NprBin}
\end{equation}
where $\mathcal{C}_p$ is the proton abundance of the cosmic rays.

If we have a sample of $N$ events corresponding to a binomial distribution with $n_0$ positive trials, an estimator
of the parameter $p$ of the binomial formula ($\mathcal{C}_{p}$ in the Eq. (\ref{NprBin})) is given by $\hat{p}=n_0/N$
and the upper, $p_{max}$, and lower, $p_{min}$, limits of the interval which contains the real value of the parameter $p$
with probability $\alpha$ (confidence level), are the solutions of,
\begin{eqnarray}
\sum^{n_{0}}_{n=0} {N \choose n} p_{max}^n (1-p_{max})^{N-n} &=& \frac{1-\alpha}{2},
\label{BinCLmax} \\
\sum^{n_{0}-1}_{n=0} {N \choose n} p_{min}^n (1-p_{min})^{N-n} &=& \frac{1+\alpha}{2},
\label{BinCLmin}
\end{eqnarray}
for $n_{0} \ne 0$ and $n_{0} \ne N$. When $n_{0}=0$ and $n_{0}=N$ the estimators of $p$ are zero and one, respectively, 
and we can obtain, at a given confidence level $\alpha$, an upper limit $p_{up}$ for the case $n_{0}=0$ and a lower limit 
$p_{low}$ for $n_{0}=N$,
\begin{eqnarray}
p_{up} &=& 1-\sqrt[N]{1-\alpha},
\label{BinCLcs} \\
p_{low} &=& \sqrt[N]{1-\alpha}.
\label{BinCLci}
\end{eqnarray}

By using the method described in subsection \ref{CompoDet} we can find an interval, $[c_1, c_2]$, corresponding 
to a given confidence level $\alpha$, for the composition of a sample. Therefore, the number of protons
of the sample, at a confidence level $\alpha$, is contained in $[n_1, n_2]=[N c_1, N c_2]$. For the case
where $n_{1} \ne 0$ and $n_{2} \ne N$ we can obtain the lower limit for the cosmic rays composition,
$\mathcal{C}_p^{min}$, by solving Eq. (\ref{BinCLmin}) with $n_0 = n_1$ and the upper limit,
$\mathcal{C}_p^{max}$, by solving Eq. (\ref{BinCLmax}) with $n_0 = n_2$. For the case in which $n_1=0$,
we just have to calculate the upper limit of the interval by solving Eq. (\ref{BinCLmax}) with $n_0 = n_2$.
The same happens for the case where $n_2=N$. We just have to calculate the lower limit of the interval
by solving Eq. (\ref{BinCLmin}) with $n_0 = n_1$. The criterion adopted to obtain the central value for
the cosmic rays composition is to take it equal to the one of the sample, $\hat{\mathcal{C}}_p = \hat{c}_p$.

Table \ref{CompoRC} shows the central values and their uncertainty at $95\%$ confidence level for the
composition of the cosmic rays inferred from the composition of the samples obtained for the points $P_i$ and 
$Q_i$ with \mbox{$i=1,2,3$} shown in table \ref{CompoInf}. As expected, the uncertainty due to the finite
size of the samples is more important for the case of $N=100$ events.
\begin{table}[h]
\begin{center}
\caption{Cosmic rays composition and its uncertainty at $95\%$ confidence level for the points $P_i$ and $Q_i$
with \mbox{$i=1,2,3$} obtained from the composition of the samples of $N=100$ and $N=1000$ events.}
\vspace{0.25cm}
\label{CompoRC}
\begin{tabular}{c c c} \hline
Points &  ${\cal C}_{p}$ for $N = 100$ &  ${\cal C}_{p}$ for $N = 1000$  \\   \hline 
$P_1$  & $0.03^{+ 0.16}_{-0.03}$ &  $0.031^{+ 0.041}_{- 0.029}$ \\ 
$P_2$  & $0.54^{+ 0.19}_{-0.20}$ & $0.537^{+ 0.064}_{- 0.068}$ \\  
$P_3$  & $0.95^{+ 0.06}_{-0.19}$ & $0.945^{+ 0.040}_{- 0.053}$ \\  
$Q_1$  & $0.18^{+ 0.19}_{- 0.14}$ &  $0.176^{+ 0.058}_{- 0.053}$ \\  
$Q_2$  & $0.47 \pm 0.20$ &  $0.469 \pm 0.060$ \\  
$Q_3$  & $0.91^{+ 0.10}_{-0.22}$ &  $0.906^{+ 0.094}_{-0.086}$ \\  \hline
\end{tabular}
\end{center}
\end{table}

\end{document}